\documentclass[sigconf,nonacm]{acmart}

\usepackage{listings}
\usepackage{xcolor}
\usepackage{booktabs}

\definecolor{codegreen}{rgb}{0,0.6,0}
\definecolor{codegray}{rgb}{0.5,0.5,0.5}
\definecolor{codepurple}{rgb}{0.58,0,0.82}
\definecolor{backcolour}{rgb}{0.95,0.95,0.92}
\definecolor{delim}{RGB}{20,105,176}
\definecolor{numb}{RGB}{106, 109, 32}
\definecolor{string}{rgb}{0.64,0.08,0.08}

\lstdefinestyle{mystyle}{
    backgroundcolor=\color{backcolour},
    commentstyle=\color{codegreen},
    keywordstyle=\color{magenta},
    numberstyle=\tiny\color{codegray},
    stringstyle=\color{codepurple},
    basicstyle=\ttfamily\footnotesize,
    breakatwhitespace=false,         
    breaklines=true,                 
    captionpos=b,
    keepspaces=true,
    numbers=left,
    numbersep=5pt,
    showspaces=false,
    showstringspaces=false,
    showtabs=false,
    tabsize=2
}

\lstdefinelanguage{json}{
    upquote=true,
    literate=
     *{0}{{{\color{numb}0}}}{1}
      {1}{{{\color{numb}1}}}{1}
      {2}{{{\color{numb}2}}}{1}
      {3}{{{\color{numb}3}}}{1}
      {4}{{{\color{numb}4}}}{1}
      {5}{{{\color{numb}5}}}{1}
      {6}{{{\color{numb}6}}}{1}
      {7}{{{\color{numb}7}}}{1}
      {8}{{{\color{numb}8}}}{1}
      {9}{{{\color{numb}9}}}{1}
      {\{}{{{\color{delim}{\{}}}}{1}
      {\}}{{{\color{delim}{\}}}}}{1}
      {[}{{{\color{delim}{[}}}}{1}
      {]}{{{\color{delim}{]}}}}{1},
}

\usepackage{seqsplit}

\newcommand*\rot{\rotatebox{90}}

\newcommand*\Mlizz{\shortstack[l]{Intel \\ i7-3770}}
\newcommand*\Mstockport{\shortstack[l]{Intel \\ Xeon 8280}}
\newcommand*\MJetson{\shortstack[l]{Jetson \\ Nano}}
\newcommand*\MRasp{\shortstack[l]{Raspberry \\ Pi 4B}}
\newcommand*\MHiFive{\shortstack[l]{HiFive \\ Unleashed}}
\newcommand*\MSpike{\shortstack[l]{Spike}}

\newcommand*\MMlizz{\shortstack[l]{Intel i7-3770}}
\newcommand*\MMstockport{\shortstack[l]{Intel Xeon 8280}}
\newcommand*\MMJetson{\shortstack[l]{Jetson Nano}}
\newcommand*\MMRasp{\shortstack[l]{Raspberry Pi 4B}}
\newcommand*\MMHiFive{\shortstack[l]{HiFive Unleashed}}
\newcommand*\MMSpike{\shortstack[l]{Spike}}

\begin{document}

\widowpenalty=10000
\clubpenalty=10000
\brokenpenalty=10000


\title{A Comprehensive and Cross-Platform Test Suite\\ for Memory Safety}
\subtitle{Towards an Open Framework for Testing Processor Hardware Supported Security Extensions}

\author{Wei~Song}
\email{songwei@iie.ac.cn}
\affiliation{%
  \institution{SKLOIS, Institute of Information Engineering, CAS \\
    School of Cyber Security, University of Chinese Academy of Sciences}
  \city{Beijing}
  \country{China}
}

\author{Jiameng~Ying}
\email{yingjiameng@iie.ac.cn}
\affiliation{%
  \institution{SKLOIS, Institute of Information Engineering, CAS \\
    School of Cyber Security, University of Chinese Academy of Sciences}
  \city{Beijing}
  \country{China}
}

\author{Sihao~Shen}
\email{shensihao@iie.ac.cn}
\affiliation{%
  \institution{SKLOIS, Institute of Information Engineering, CAS \\
    School of Cyber Security, University of Chinese Academy of Sciences}
  \city{Beijing}
  \country{China}
}

\author{Boya~Li}
\email{liboya@iie.ac.cn}
\affiliation{%
  \institution{SKLOIS, Institute of Information Engineering, CAS \\
    School of Cyber Security, University of Chinese Academy of Sciences}
  \country{China}
}

\author{Hao~Ma}
\email{mahao@iie.ac.cn}
\affiliation{%
  \institution{SKLOIS, Institute of Information Engineering, CAS \\
    School of Cyber Security, University of Chinese Academy of Sciences}
  \city{Beijing}
  \country{China}
}

\author{Peng Liu}
\email{pxl20@psu.edu}
\affiliation{%
  \institution{The Pennsylvania State University}
  \city{University Park}
  \country{USA}
}

\begin{abstract}

  Memory safety remains a critical and widely violated property in reality.
  Numerous defense techniques have been proposed and developed
  but most of them are not applied or enabled by default in production-ready environment
  due to their substantial running cost.
  The situation might change in the near future because
  the hardware supported defenses against these attacks
  are finally beginning to be adopted by commercial processors, operating systems and compilers.
  We then face a question as
  there is currently no suitable test suite to measure the memory safety extensions supported on different processors.
  In fact, the issue is not constrained only for memory safety but all aspect of processor security.
  All of the existing test suites related to processor security lack some of the key properties,
  such as comprehensiveness, distinguishability and portability.
  
  As an initial step,
  we propose an expandable test framework for measuring the processor security
  and open source a memory safety test suite utilizing this framework.
  The framework is deliberately designed to be flexible so it can be gradually extended to
  all types of hardware supported security extensions in processors.
  The initial test suite for memory safety currently contains 160 test cases covering
  spatial and temporal safety of memory, memory access control,
  pointer integrity and control-flow integrity.
  Each type of vulnerabilities and their related defenses
  have been individually evaluated by one or more test cases.
  The test suite has been ported to three different instruction set architectures (ISAs)
  and experimented on six different platforms.
  We have also utilized the test suite to explore
  the security benefits of applying different sets of compiler flags available
  on the latest GNU GCC and LLVM compilers.
  

\end{abstract}

\maketitle

\section{Introduction}

It is widely known that the lack of memory safety in low-level languages like C/C++
enables attackers to maliciously access data in the memory,
alter the value of key variables such as code pointers,
hijack the behavior of applications,
or even take full control of a computer system~\cite{Szekeres2013}.
In addition to the endless endeavor to discover and then fix the individual bugs
scattered around the enormous software ecosystem,
researchers have proposed a multitude of defense techniques in recent decades
to thwart one or multiple types of attacks exploiting memory bugs.
However, memory safety remains a critical and widely violated property in reality
as most of the operating systems (OSes) and applications running on the off-the-shelf computers
are not adequately defended.

Some of the early defense techniques,
such as stack canary~\cite{Cowan2003}, data execution prevention (DEP) and address space layout randomization (ASLR)~\cite{PaX2001},
have been widely adopted by commercial processors, operating systems (OSes) and compilers,
which in turn triggers the evolution of the more complicated and evasive variants of attacks,
such as return-oriented programming (ROP)~\cite{Shacham2007, Tran2011}, jump-oriented programming (JOP)~\cite{Bletsch2011},
counterfeit object-oriented programming (COOP)~\cite{Schuster2015} and data-oriented programming (DOP)~\cite{Chen2005, Hu2016}.
To battle with these new attacks, numerous defense techniques have been proposed and developed.
Some of the extensively studied techniques include
type enforcement by tagging the pointers~\cite{Necula2005, Arora2006a, Kwon2013, Dhawan2014, Woodruff2014},
pointer authentication (PA)~\cite{Cowan2003a, Mashtizadeh2015}, code-pointer integrity (CPI)~\cite{Kuznetsov2014},
control-flow integrity (CFI)~\cite{Abadi2009, Burow2017} and data-flow tracking~\cite{Chang2008, Song2016}.
Yet most of them are not applied or enabled by default in production-ready environment due to their substantial running cost.
For example, CFI has been implemented in LLVM and GCC~\cite{Tice2014} but it is almost disabled by default.
The support of Intel MPX (memory protection extension)~\cite{Ramakesavan2016} was added but later dropped by major compilers.
As a result, even naive buffer overflow bugs can be exploited without being detected.

On the bright side, the situation might change in the near future,
since the hardware supported defenses against the more complicated and evasive variants of attacks
are finally beginning to land on commercial processors, OSes and compilers.
The Intel control-flow enforcement technology (CET)~\cite{CET2019}, initially drafted in 2016 to thwart ROP and JOP related attacks,
is included in the 11th generation Tiger Lake processors (expected in the middle of 2021).
It will be supported by Windows 10 through the hardware-enforced stack protection.
The Arm pointer authentication (PA) was added in the ARMv8.3-A~\cite{Brash2016}
to facilitate defenses like CPI.
It has already been implemented in the Apple A-series processors after A12
and supported in LLVM from version 8~\cite{PALLVM2021}.
The Arm memory tagging extension (MTE) was proposed in ARMv8.5-A~\cite{Bannister2019}
to further enhance memory safety in general
and Google has announced its willingness to support it in Android~\cite{Serebryany2019}.

With no doubt, hardware supported security extensions would significantly reduce the cost of applying defenses
and gradually spread to all computer architectures and processor implementations.
This tide of changes
would bring us a new set of questions:
\begin{itemize}
\item \emph{Q1: How to compare the security provided by different processors?}
  When picking a proper processor (actually a minimal platform comprising of a processor, an OS and a compiler)
  for a specific system to be developed,
  a system designer may desire to choose the one providing the maximum security protection
  from all candidate processors within the performance and cost budget.
\item \emph{Q2: How to choose the right compiler and compiler flags (OS features)
  for the best security-performance trade-off?}
  When shipping an application to a target processor,
  an application designer might wish to apply a proper set of compiler flags on a suitable compiler
  that enables the most security extensions available on the target processor
  while maintaining the performance target.
\item \emph{Q3: How to evaluate the security benefit of a new defense on a certain processor?}
  As security researchers, 
  we would also like to systematically evaluate the security benefit of a newly proposed defense on a target processor.
\end{itemize}

We believe the answer is to create a test suite
which is comprehensive enough to quantitively evaluate the safety of a processor
while also portable enough to comparatively evaluate multiple processors and be expanded to evaluate new defenses, vulnerabilities and architectures.
Obviously, this will be an extremely challenging while long-lasting work.

As an initial step,
we propose an expandable test framework for measuring the processor security
and open source a memory safety test suite utilizing this framework.
The framework is deliberately designed to be flexible so it can be gradually extended to
all types of security features supported by the processor architecture and hardware (micro-architecture),
such as defenses targeting the cache side-channels and speculative execution attacks.

The initial test suite for memory safety currently contains 160 test cases covering spatial and temporal safety of memory, memory access control,
pointer integrity and control-flow integrity.
Each type of vulnerabilities and their related defenses
have been individually evaluated by one or more test cases.
The test suite has been ported to three different instruction set architectures (ISAs)
and experimented on six different platforms.
We have also utilized the test suite to explore
the security benefits of applying different sets of compiler flags available on the latest GNU GCC and LLVM compilers.

The rest of this paper is organized as follows:
Section~\ref{sec:motivation} explains the motivation behind this work.
Section~\ref{sec:threat} describes the assumptions made in this paper.
Section~\ref{sec:frame} illustrates the testing framework proposed to construct and run the test suite.
The available test cases are summarized in Section~\ref{sec:test-case}
with the implementation challenges illustrated in Section~\ref{sec:challenge}.
Section~\ref{sec:result} analyzes the test results on the six ported platforms.
Section~\ref{sec:discuss} discusses the related and future work.
Finally, the paper is concluded by Section~\ref{sec:con}.

\section{Motivation}\label{sec:motivation}

This work is motivated by the lack of testing for processor security in both industry and academia.

\subsection{The Lack of Testing in Industry}

The systematic testing for processor security is almost non-existing in industry.
Major processor manufacturers and OS vendors may have internal tests but they are not in any way shared or publicized.
They normally sit on the back seat and are passively driven to respond to vulnerabilities
only when they are discovered and even already publicized.
Take the discovery of Meltdown and Spectre for example,
it was reported that Intel did not inform the U.S. National Security Agency until they were made public~\cite{Nellis2018}.
Microsoft eventually mobilized hundreds of people across the company and urgently brought in external experts in response
and negotiated an extension of the disclosure date to 120 days~\cite{Fogh2018}.
All of these indicate that the transient execution vulnerabilities have struck the whole industry off guard.
If processor designs and the manufactured central processing units (CPUs) were systematically and regularly tested for hardware security,
the situation might be significantly improved,
at least the response would be much more prompt.
However, there is no evidence of such testing facility being proposed or developed.
The mitigation of processor vulnerabilities are still internally evaluated, responded and potentially resolved
by individual processor manufacturers.
This would leave the smaller players, who cannot spare engineers to respond to these security concerns
or simply are not knowledgeable enough to handle these concerns, out in the insecure wild along with their customers.

The situation on the customer side is even more stark.
End customers are forced to unconditionally trust the security claims from processor manufacturer without any means to verify them.
Even governments may have very limited ways to measure the security of the processors deployed on the safety critical infrastructures.

If there is a generic testing framework capable of measuring the processor security,
even if the measurement covers only the publicized hardware vulnerabilities,
processor manufacturers may benefit from it by obtaining an early warning on the potentially unprotected vulnerabilities
and the customer can earn extra confidence on the security claims produced by the manufacturers,
as well as a mean to compare the security levels of different CPUs.

\subsection{Insufficient Testing in Academic Research}

The research in academia is normally more advanced and open than those in industry,
which is the same for the security evaluation for processors.
However, the existing research is still insufficient.

Some test suites have already been proposed in the literature. 
The Juliet C/C++ and Java test suite~\cite{Boland2012} is a widely used collection of flawed programs
to test the effectiveness of static analyzers in compilers.
RIPE (runtime intrusion prevention evaluator)~\cite{Wilander2011} provides a combination of synthetic buffer overflow attacks
as a way to test the effectiveness of related defenses.
RIPE is perhaps the most relevant and well-known test suite for processor security and
has already been extensively used in architectural researches to verify their defenses against control-flow attacks~\cite{Kuznetsov2014,Song2016},
especially the ROP variants.
Each test case in RIPE is a relatively complete attack specified by five factors:
the location of the buffer to overflow,
the type of code pointer to corrupt,
the type of overflow attack,
the type of shell code
and the function to be abused in the overflow attack.
Since each factor has multiple choices,
RIPE contains as many as 850 test cases to cover the ROP variants of control flow alteration caused by buffer overflow.
It is obvious that this way of testing (by relatively complete attacks involving multiple attack factors) and reaching coverage (by exhausted testing)
is very hard to be extended into a full-fledged suite for all aspects related to processor security.

None of the existing test suites can, or be expanded to, conduct a thorough measurement of the safety of a processor
due to the lack of some of the following key properties:
\begin{itemize}
\item \emph{Comprehensiveness}:
  The test suite should have a wide coverage of all types of vulnerabilities and defense techniques
  especially those potentially supported by the processor hardware.
  Most of existing test suites either cover only a portion of the vulnerabilities and protections, such as RIPE,
  or concentrate on only the software defenses, such as Juliet.
\item \emph{Distinguishability}:
  To provide a quantitative evaluation, the test suite should provide per-vulnerability and per-defense test cases
  which concentrate on a single attack factor (such as one type of target or one attack technique).
  Such level of distinguishability is not available in RIPE.
\item \emph{Structure}:
  The test suite should be produced in a structured way rather than a collection of ad-hoc test for individual vulnerabilities or defenses.
  This structure is crucial for the test suite to reach a coverage which is understandable and reasonable.
\item \emph{Portability}:
  The test suite should be able to be ported to multiple platforms using various computer architectures,
  OSes and compilers with a minimum effort.
  Once ported to a certain type of platforms,
  the test suite should run out-of-the-box on all variants of the same platform.
  When a new type of attacks or protection emerges,
  individual test cases should be able to be added with a minimum effort.
  None of the existing test suites maintain good portability.
\end{itemize}

In this work, we seek to provide these key properties in a expandable testing framework by the following method.
\begin{itemize}
\item \emph{Comprehensiveness}:
  It is impossible to bring up a comprehensive test suite for all aspects of processor security in a short period of time
  but we can gradually grow the aspects covered by the test suite.
  As in the initial test suite, major aspects of the memory safety are covered.
\item \emph{Distinguishability}:
  Every test case in the test suite is made concentrating on a single type of vulnerabilities or defenses.
  The runtime context of individual test cases are created by a rather brutal forced way, as shown in Section~\ref{subsec:con-tc},
  to deliberately avoiding introducing extra vulnerabilities.
  To some extent, each test case is a partial attack that is normally executed as a single step in a relatively complete attack.
\item \emph{Structure}:
  The dependence of individual types of vulnerabilities and defenses are analyzed and recorded using a relation graph,
  as described in Section~\ref{subsec:coverage}.
  By utilizing this graph, test cases can collect crucial pre-knowledge from the results of other tests
  and use it to improve test accuracy.
  The order of testing and a reasonable coverage are also derived form this relation graph.
\item \emph{Portability}:
  The code of individual test cases are divided into platform independent and dependent parts.
  All the platform dependent parts are moved to a shared platform-specific library as described in Section~\ref{subsec:con-tc}.
  This significantly improves the portability of the test suite.
\end{itemize}

\section{Threat Model}\label{sec:threat}

\emph{Adversarial capability}:
We assume unprivileged attackers with the ability to execute and control the input of user programs on an OS.
It is also assumed that user programs commonly contain memory safety vulnerabilities
which might be exploited to achieve arbitrary reads and writes into the program address space.
As a consequence,
any attacks utilizing the existing memory safety vulnerabilities to
leak information, corrupt data, and hijack control flow are in scope.
We limit the current test suite to attacks utilizing the existing memory safety vulnerabilities
to attack the user level data space.
Therefore, attacks that using side-channels, such as cache side-channels~\cite{Yarom2014, Yan2017}
and transient execution attacks~\cite{Canella2019},
and attacks targeting the kernel space
are out of the scope of this paper
but, of course, they are the targed attacks for the future expended version of this test suite.

\emph{Definition of a platform}:
The safety of a processor is measured by running the test suite on a minimum running environment,
namely a platform,
comprising the processor under test and a minimum OS running on it.
This minimum OS includes a kernel and a small number of runtime libraries
just enough for the execution of the user programs potentially under attack.
All user programs are assumed to be compiled by a common compiler using similar compiler flags.

\emph{Hardening assumptions}:
We assume some memory safety protection features have been implemented
by either software or hardware
but the exact details of these features may not make available to the test suite
unless they are visible to user-level programs (through environmental variables)
or indicated by compiler flags fed to the test suite.
If a safety protection feature is implemented,
it should have already been supported by the OS and the compiler;
therefore, the protection is enabled by default or can be switched on by compiler flags. 

\section{The Framework of Testing}\label{sec:frame}

This section proposes a flexible testing framework suitable for measuring the security provided by a processor.
While in the description, we use memory safety as the targeted aspect of security for measurement.

\subsection{Scope of Testing}\label{subsec:scope}

Memory safety could be viewed as a set of memory safety properties.
All memory corruption vulnerabilities and exploitation techniques rely on the absence of certain memory checks on one or more properties
which allow values in memory to be maliciously leaked or altered against their original meaning in the program.

For example, a buffer overflow attack happens when the action of modifying the value of a buffer goes beyond its boundary.
Modifying beyond the buffer boundary is the memory safety property that has been violated due to the lack of a proper boundary check.
The fact that the modifying of the buffer can be made beyond the boundary is a vulnerability.
To eliminate this vulnerability, the specific buffer can be patched with the missing boundary check in the source code,
or this check can be universally added to all buffers to prevent them from being exploited by the same type of vulnerabilities.
We consider the latter as a defense.
A defense can be implemented purely by software but with significant slowdown~\cite{Nagarakatte2009, Serebryany2012},
or it can be enforced by hardware through memory tagging or fat pointers
while with the help of compilers and libraries~\cite{Nagarakatte2012}.
We consider the former as a pure software defense
while the latter as a hardware supported security enhancement.

Memory safety is not the responsibility of the processor hardware but the whole system, including the programming language,
the compiler, the operating system, the runtime libraries, the architecture and finally the processor hardware.
Not all memory safety properties are enforceable by the processor hardware
but when they are, adopting the hardware supported enforcement normally brings significant benefit in performance.
The objective of the proposed test suite is to measure the security boost
brought by the security enforcement potentially implemented in the processor under test.
For this reason, the aforementioned buffer overflow attack should be tested
because the missing boundary check can be enforced by the processor hardware through memory tagging or fat pointers.
In other words,
\emph{if the absence of a memory check can be exploited by attackers
  and the check can be efficiently enforced by the processor hardware (and architecture) with (or even without) the help of software,
  this check is considered a potential hardware supported security enhancement
  that should be covered by the test suite.}

Although the concept might be easy to digest, the reality is still very much complicated.
At least for memory safety, most architectural security enhancements require software (especially the compilers) involvement.
The test suite aims to discover whether certain checks have been enforced with the support from the processor hardware;
however, the fact that a test suite must be executed in a software environment
blurs the boundary between software and hardware.
A test might find the existence of a check but it cannot simply tell whether it is hardware supported or purely software enforced.
As a result, \emph{the test suite should be considered as a best-effort approach to measure the memory safety supported by the processor hardware.}
The test suite should be compiled by compilers pre-installed in the production-ready OS
and run on the OS with a minimal execution environment where only the necessary libraries are installed.
In this scenario, checks (defenses) found by the test suite
are likely to be enforced with the support from processor hardware (or at least the essential runtime library)
rather than some third-party software.

With all the above in mind, we consider
\emph{the processor hardware,
the ISA, the kernel of the production-ready OS,
the default compiler,
and the essential runtime libraries}
as the scope of testing.
This leaves the security features available in some binary-level translation and execution tools,
and kernel patches (such as Grsecurity{\textsuperscript{\textregistered}}~\cite{OSS2021}) out of the scope.
Defenses that are usually implemented and enforced purely by software are not deliberately evaluated
while each hardware enforceable memory check (defense) should be covered by a specific test case.
Similarly, the test suite does not verify the existing software running on the target platform,
including the kernel and the runtime libraries,
is free of vulnerabilities.
In fact, it assumes all programs may contain vulnerabilities
but they would become difficult to exploit when proper defenses are enforced.

\subsection{Definition of Coverage}\label{subsec:coverage}


Existing test suites choose to 
launch complete attacks exploiting 
a number of vulnerabilities
and utilizes multiple attack techniques at different stages~\cite{Poe2006, Wilander2011}.
If a coverage is defined as all possible ways of attacks,
it would need to cover all the potential combinations of vulnerabilities and attack techniques,
which becomes a task impossible to fulfill.
As a result, existing test suites normally cover only a set of related types of vulnerabilities,
leaving a significant amount of attack variants uncovered~\cite{Li2020}.

Instead of trying to reach a partial coverage by launching complete attacks,
this test suite tries to cover a wide range of vulnerabilities
with a large number of small and concentrated test cases.
Each test cases endeavors to check the existence of a single type of memory vulnerabilities
or the absence of a specific memory check. 
By doing this, the test suite satisfies the requirement for the first two key properties,
\emph{comprehensiveness} and \emph{distinguishability},
as described in Section~\ref{sec:motivation}.

Since each test case concentrates on a single type of vulnerabilities,
which is normally exploited as a step in a complex attack,
test cases are not mutually isolated but rather rely on each other to provide key information.
We try to describe this relationship between vulnerabilities by a graph,
which then helps the test suite execute all test cases with a proper order and reach a coverage faster.
More importantly, we can derive a clear definition of coverage based on this graph.
Three types of relations have been considered: 
\begin{itemize}
\item \emph{Dependency}:
  If a type of vulnerabilities \emph{A} can be exploited only when another type \emph{B} is exploitable,
  \emph{A} depends on \emph{B}.
  Take the classic ROP attack as an example.
  To alter the return address stored on the stack and make it point to a code gadget,
  the attacker needs to know the address of the gadget,
  which usually depends on that some code pages are readable as ASLR is enforced.
  In this case, \emph{A: alteration of return address} depends on the \emph{B: readability of code pages}.
\item \emph{Specialization}:
  If a constraint is applied on a type of vulnerabilities \emph{B}, which creates a new type of vulnerabilities \emph{A},
  \emph{A} is a special case of \emph{B}.
  Let us use the same ROP example.
  Since the alteration of return address happens on the stack,
  it can be considered as applying a type constraint (code pointer) on a general stack modification vulnerability.
  Therefore, \emph{A: alteration of a return address} is a special case of \emph{B: data alteration on the stack}.
\item \emph{Relaxation}:
  If two types of vulnerabilities \emph{A} and \emph{B} are similar
  but \emph{B} is more evasive than \emph{A} in the view of memory checks,
  \emph{A} is a relaxed form of \emph{B}.
  This type of relation could be hard to conceive but there are abundant examples in practice.
  Still considering the ROP attack,
  replacing the return address with a wrong call site address is a more evasive attack
  than replacing it with a non-call-site address.
  In this case, \emph{A: ROP using a non-call-site address}
  is a relaxed form of \emph{B: ROP using a wrong call site address.}
\end{itemize}

In all types of relation,
if \emph{B} is found eliminated by a memory check,
\emph{A} is almost certainly eliminated by the same memory check
and is therefore unnecessary to be tested.
In this scenario, we call \emph{B} the prerequisite of \emph{A}.

\begin{figure}[bt]
\centering{
\includegraphics[width=0.35\textwidth]{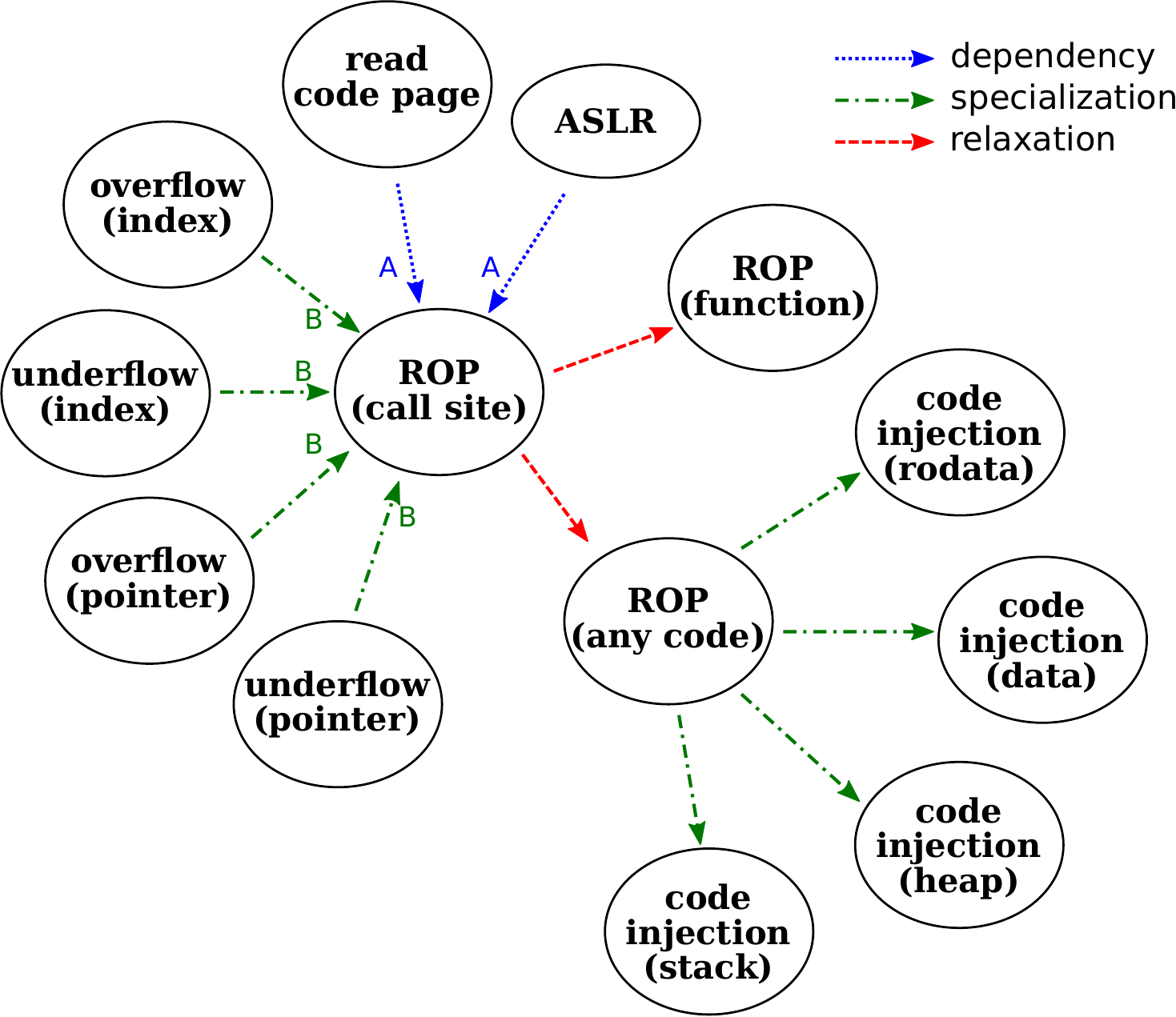}
}
\caption{Relation graph related to ROP attacks}
\label{fig:relation-graph}
\end{figure}

The prerequisites of all test cases can be depicted by a labeled directed graph,
which is called a relation graph in this paper.
The one describing the ROP related vulnerabilities 
is shown in \figurename~\ref{fig:relation-graph}.
Each node denotes a test case targeting one specific type of vulnerabilities or defenses
while an arc between two nodes describes the prerequisite relation.
The three types of relations are depicted in different colors and dash lines.
To hijack a return address,
the attacker needs to search for a proper gadget in the code section.
This could be done by either using an offset collected from static binary analysis when ASLR is not enabled
or analyzing code pages at runtime~\cite{Snow2013}.
Therefore, \textbf{ROP (call site)} \emph{depends} on the missing of \textbf{ASLR} or the possibility to \textbf{read code page} at runtime.
Since the return address is a special variable stored on the stack,
altering it is a \emph{special case} of stack overflow attacks.
Consequently, \textbf{ROP (call site)} takes four \textbf{over/underflow} tests as prerequisites as well.
Three ROP tests are included in the relation graph:
returning to a \textbf{call site}, using a \textbf{function} as a gadget, and returning to \textbf{any code}.
Since \textbf{ROP (call site)} is the most evasive of the three,
we consider \textbf{ROP (function)} and \textbf{ROP (any code)} as two \emph{relaxed} cases.
Finally, the gadgets could be codes directly injected in read-only data,
global data, heap and stack.
Depending on the location,
we have four \emph{special cases} of \textbf{ROP (any code)}:
\textbf{code injection} in \textbf{rodata}, \textbf{data}, \textbf{heap} and \textbf{stack}.

The prerequisites for \textbf{ROP (call site)} is complicated.
It has six prerequisites but it is enabled when
\emph{one of} \textbf{ASLR} and \textbf{read code page} passes, \textsc{and} \emph{one of} the four \textbf{over/underflow} tests passes.
This is literally a two-level relationship:
Prerequisites are divided into subsets.
In each subset, the prerequisite is met if \emph{one of} the test cases passes.
The final prerequisite is met when the results of all subsets are \emph{AND}ed together and turns to be true.
To illustrate such a two-level relationship in the relation graph,
the arcs of each subset are labeled by a unique letter.
Labels are omitted when only one level of relation is needed. 

According to the prerequisites defined in the relation graph,
the test suite can avoid testing both \textbf{ROP (function)} and \textbf{ROP (any code)}
if \textbf{ROP (call site)} fails.
The same is for the four \textbf{code injection} tests if \textbf{ROP (any code)} fails.
\emph{
  Assuming a relation graph has exhausted all types of vulnerabilities and defenses related to certain aspects of memory safety,
  reaching a full coverage is equivalent to obtain a reachability graph
  by starting from all nodes without input arc.
}
To this end, the test suite satisfies the \emph{structure} requirement described in Section~\ref{sec:motivation}.

We believe the concepts of relation graph and the derived coverage applies to all aspects of processor security
beyond just the memory safety.
For example, a conflict-based cache side-channel attack\cite{Yan2017} is a multi-step attack comprising at least 
the construction of eviction sets, priming the target cache sets, triggering the victim rerun and finally probing the cache.
The success of each step may depend on some pre-knowledge of the micro-architecture and
certain ways of utilizing the micro-architectural features.
Hardware supported defenses in processors may thwart some of these steps,
such as increasing the difficulties in constructing eviction sets using randomized caches~\cite{Qureshi2019, Werner2019}
or effectively prohibiting priming by enforcing cache partitioning~\cite{Liu2016}.
All of these attacking steps, defenses, and micro-architectural features (potential vulnerabilities)
can be described using the relation graph and tested accordingly using a set of concentrated test cases.
In other words, the testing framework can be expanded to all aspects of processor security.

Another important issue related to relation graph is its construction and correctness.
Currently the relation graph is hand crafted by meanually building up the dependence between different
vulnerabilities, defenses and related techniques according to our own understanding of the related literature.
Unavoidably this hand crafted relation graph contains human errors due to our limited knowledge
and potential misundertanding.
However, the test suite is opensourced (GitHub repo is provided at the end of this paper)
and the relation graph is recorded in a JSON file (as described by the next paragraph)
included in the test suite.
We will gradually evlove the relation graph along with the expansion of the test suite
and our improved understanding of the related issues.
We also hope the test suite would become a community project which attracts contribution from the whole community
including the corrections on the relation graph.

To utilize the prerequisites at runtime,
the relation graph is stored in a configuration JSON file
and analyzed at runtime by a simple test scheduler written in python.
Listing~\ref{list:configure-require} demonstrates a part of the configuration file describing \figurename~\ref{fig:relation-graph}.
Every record is a test case depicted as a node in the relation graph,
with the name mapping between JSON records and nodes in the relation graph described in Table~\ref{tab:name-map}.
The prerequisites are listed in the ``require'' property,
which is a list of lists.
Each inner list contains prerequisites within the same subset
while results of all inner lists of are \emph{AND}ed together to
decide whether the case is tested.

\begin{figure}[bt]
\centering{
\begin{minipage}{0.43\textwidth}
\begin{lstlisting}[language=json, caption=\texttt{configure.json}, label=list:configure-require]
{
  ... // omit some code

  "return-to-wrong-call-site": {
    "require": [ [ "read-func", check-ASLR"],
                 [ "overflow-write-index-stack",
                   "underflow-write-index-stack",
                   "overflow-write-ptr-stack",
                   "underflow-write-ptr-stack" ] ]
  },

  "return-to-non-call-site": {
    "require": [ [ "return-to-wrong-call-site" ] ]
  },

  "return-to-func": {
    "require": [ [ "return-to-wrong-call-site" ] ]
  },

  "return-to-instruction-in-rodata": {
    "require": [ [ "return-to-non-call-site" ] ]
  },

  "return-to-instruction-in-data": {
    "require": [ [ "return-to-non-call-site" ] ]
  },

  "return-to-instruction-in-heap": {
    "require": [ [ "return-to-non-call-site" ] ]
  },
  
  "return-to-instruction-in-stack": {
    "require": [ [ "return-to-non-call-site" ] ]
  },

  ... // omit some code
}
\end{lstlisting}
\end{minipage}
}
\end{figure}

\begin{table}[bt]
\footnotesize
\begin{center}
  \caption{Name mapping between JASON and \figurename~\ref{fig:relation-graph}}\label{tab:name-map}
  \begin{tabular}{ll}
    \toprule
    Node Name    & JSON record name          \\
    \midrule
    read code page & \texttt{read-func} \\
    ASLR           & \texttt{check-ASLR} \\
    overflow (index) & \texttt{overflow-write-index-stack} \\
    underflow (index) & \texttt{underflow-write-index-stack} \\
    overflow (pointer) & \texttt{overflow-write-ptr-stack} \\
    underflow (pointer) & \texttt{underflow-write-ptr-stack} \\
    ROP (call site) & \texttt{return-to-wrong-call-site} \\
    ROP (any code) & \texttt{return-to-non-call-site} \\
    ROP (function) & \texttt{return-to-func} \\
    code injection (rodata) & \texttt{return-to-instruction-in-rodata} \\
    code injection (data) & \texttt{return-to-instruction-in-data} \\
    code injection (heap) & \texttt{return-to-instruction-in-heap} \\
    code injection (stack) & \texttt{return-to-instruction-in-stack} \\
    \bottomrule
  \end{tabular}
\end{center}
\end{table}

\subsection{Construction of Test Cases}\label{subsec:con-tc}

\emph{Portability} is one of the major concerns in the construction of test cases.
There is no doubt that new types of vulnerabilities will be discovered in the future,
along with new defense techniques.
The test suite tries to cover all the known types of vulnerabilities
while allowing new test cases to be easily added. 
The ease by which new test cases can be added is then crucial for the usefulness of the test suite.
In addition, new computer architectures and ISA extensions will be constantly proposed.
Instead of painstakingly re-implement all test cases whenever a new ISA is added or a new OS is supported,
the test suite should try to reuse the platform-independent part 
while limiting the platform-specific part to relatively small code segments.
For these reasons, the test suite is composed of two parts:
\emph{individual platform-independent test cases}
and \emph{a shared platform-specific support library}.

Each test case is written in a short (and usually malformed) C++ program
trying to test a single type of vulnerabilities.
If the vulnerabilities exist, the test case should exploit the vulnerability and return zero;
otherwise, a non-zero exit code is used to indicate when and where the test fails or crashes.
Whenever possible,
the related vulnerabilities involved in each test cases is minimized by
directly creating a vulnerable context at runtime.
Most code is written in a portable and compiler-independent manner.
However, when a (usually malicious) behavior might be detected and disturbed by a compiler, 
extra code might be added to prevent the compiler optimization
and snippets of assembly code might be used to replace the C++ code
as most compilers would not analyze the embedded assembly, let alone optimize it.
Since the added assembly code snippets are unavoidably platform-specific,
they are extracted from test cases and moved to the shared support library.

For example, Listing~\ref{list:test-case} illustrates the test for \texttt{\seqsplit{return-to-wrong-call-site}}
which hijacks the return address to a valid call site within static analysis. 
The \texttt{main()} function is a bare minimum.
It first reads \texttt{offset} from input as it is later needed by \texttt{helper()}.
The value of this \texttt{offset} is actually selected by the test scheduler,
which will be explained in details in Section~\ref{subsec:runtime-var}.
The \texttt{main()} function then calls \texttt{helper()} twice (line 17 and 23)
with the latter one trying to alter its own return address.
If the \texttt{helper()} function succeeds, it would return to line 19 labeled by \texttt{main\_mid},\footnote{
  An assembly label is used because C++ labels are invisible outside the function body.
} and exit with a 0.
Otherwise, the program may exit with a non-zero code 2,
indicating the test fails to override the return address.
The exit code is initialized to 0 and stored in a volatile global variable \texttt{grv}
to escape compiler optimization.

The \texttt{helper()} function is also straightforward.
\texttt{FORCE\_NOINLINE} stops the function from being inlined using a compiler attribute.
\texttt{ENFORCE\_NON\_LEAF\_FUNC} inserts some code to guarantee that
\texttt{helper()} is not a leaf function and the return address is always pushed on the stack
even on RISC (reduced instruction set computer) machines.
The embedded assembly \texttt{MOD\_STACK\_LABEL(main\_mid, offset)} replaces the return address with the address of \texttt{main\_mid}
when the function is called the second time (\texttt{grv == 2}).
The location of the return address on the stack is pinpointed by the program input \texttt{offset},
which is the global parameter collected by the \texttt{main()} function indicating the offset to the stack pointer.
As the value of \texttt{offset} varies on different platforms or even with different compiler flags.
This is why
it is detected by other tests and then provided by the test scheduler
(see Section~\ref{subsec:runtime-var} for more details).
Finally, \texttt{helper()} revises the exit code to 0 just before return.
If the return address is successfully modified and \texttt{helper()} indeed returns to \texttt{main\_mid},
the program would exit as normal (code 0) in the \texttt{main()} function.\footnote{
  We use \texttt{exit()} (from \texttt{cstdlib}) here as the call stack is corrupted.
}

\begin{figure}[bt]
\centering{
\begin{minipage}{0.43\textwidth}
\begin{lstlisting}[language=C++, caption=\texttt{return-to-wrong-call-site.cpp}, label=list:test-case]
#include <cstdlib>
#include "include/assembly.hpp"

static volatile int grv = 0;
int offset; // offset to stack pointer  

void FORCE_NOINLINE helper() {
  ENFORCE_NON_LEAF_FUNC;
  if(++grv == 2) {
    MOD_STACK_LABEL(main_mid, offset);
    grv = 0;
  }
}

int main(int argc, char* argv[]) {
  offset = 8 * (argv[1][0] - '0');
  helper();

  DECL_LABEL(main_mid);
  if(grv == 0)
    exit(grv);

  helper(); // hijack return address
  exit(grv);
}
\end{lstlisting}
\end{minipage}
}
\end{figure}

Most platform-specific codes are written in embedded assembly and referenced by macros,
which are defined in the shared support library.
As shown in Listing~\ref{list:shared-assembly},
the header file included in test cases (\texttt{include/assembly.hpp}) is just a wrapper
pointing to the correct header prepared for the target platform.
Assuming the target platform is a x86-64 machine,
macro \texttt{\_\_x86\_64} is predefined by most compilers
and \texttt{x86\_64/assembly.hpp} is then included by detecting this predefined macro.

\begin{figure}[bt]
\centering{
\begin{minipage}{0.43\textwidth}
\begin{lstlisting}[language=C++, caption=\texttt{include/assembly.hpp}, label=list:shared-assembly]
... // omit some code

#ifdef __x86_64
  #include "x86_64/assembly.hpp"
#endif

... // omit some code
\end{lstlisting}
\end{minipage}
}
\end{figure}

Listing~\ref{list:x86-assembly} reveals parts of support library for the x86-64 architecture (\texttt{x86\_64/assembly.hpp}). 
\texttt{FORCE\_NOINLINE} is short for a builtin attribute for non-inline functions.
To enforce the non-leaf condition, \texttt{ENFORCE\_NON\_LEAF\_FUNC} inserts an irremovable external call.
\texttt{MOD\_STACK\_LABEL(label, offset)} is used to replace the return address stored at \texttt{[\$rsp+offset]} with
the address of \texttt{label}
in a rather brutal forced manner.
This avoids the use of traditional buffer overflow attacks and
probably escapes from most compile-time detection,
and allows the test case to concentrate on the modification of return address to a non-call-site vulnerability itself.

\begin{figure}[bt]
\centering{
\begin{minipage}{0.43\textwidth}
\begin{lstlisting}[language=C++, caption=\texttt{x86\_64/assembly.hpp}, label=list:x86-assembly]
... // omit some code

#define FORCE_NOINLINE __attribute__((noinline))

extern int dummy_leaf_rv;
extern int dummy_leaf_func(int);
#define ENFORCE_NON_LEAF_FUNC \
  dummy_leaf_rv = dummy_leaf_func(dummy_leaf_rv);

// declare a label in assembly
#define DECL_LABEL(label) asm volatile(#label ":")

// modify stack
#define MOD_STACK_LABEL(label, offset)        \
  asm volatile(                               \
    "lea " #label "(%%rip), %%rax;"           \
    "movl %0, %%ecx;"                         \
    "movslq %%ecx, %%rcx;"                    \
    "addq %%rsp, %%rcx;"                      \
    "movq %%rax, (%%rcx);"                    \
    : : "r"(offset) : "rax", "rcx"            )

... // omit some code
\end{lstlisting}
\end{minipage}
}
\end{figure}

Assuming the functions of the macros defined in the shared support library are modular enough,
supporting a new platform should be an easy task.
A new ``\texttt{assembly.hpp}'' header file should be created to
define all the platform-specific macros and functions using the new architecture.
Currently there are only around 20 macros that are needed to be ported for each new architecture.
This new header file is then added to the common header file \texttt{include/assembly.hpp}
while checked by a predefined macro unique to the new platform.
Adding a new test case for a newly discovered type of vulnerabilities
may involve three steps:
(1) Implement the test case using the shared support library.
(2) If new macros are needed, they should be added to all supported platforms.
(3) A record is added in the configuration file.

\section{Summary of Test Cases}\label{sec:test-case}

This section summerize the individual test cases provided by the initial test suite for memory safety.

\subsection{Spatial Safety}

Spatial safety refers to the property that memory accesses are always in compliance with
the proper data boundaries and the scope of visibility defined by the program.
Any access outside the boundary or the scope of visibility is considered insecure.
It is violated by classic buffer overflow attacks~\cite{Spafford1989},
which is then used to smash the stack and alter the stored return address~\cite{AlephOne1996}.
Heap attacks also resort to buffer overflow to access/alter data across the object boundary~\cite{Conover1999}.
Currently, the test suite has 98 test cases to measure spatial safety.

\subsubsection{General read out-of-boundary}
16 test cases in total.\\
\begin{footnotesize}\begin{tabular}{|l|}
\midrule
\texttt{overflow-read-index-stack} \\
\texttt{overflow-read-index-heap} \\
\texttt{overflow-read-index-data} \\
\texttt{overflow-read-index-rodata} \\
\texttt{overflow-read-ptr-stack} \\
\texttt{overflow-read-ptr-heap} \\
\texttt{overflow-read-ptr-data} \\
\texttt{overflow-read-ptr-rodata} \\
\texttt{underflow-read-index-stack} \\
\texttt{underflow-read-index-heap} \\
\texttt{underflow-read-index-data} \\
\texttt{underflow-read-index-rodata} \\
\texttt{underflow-read-ptr-stack} \\
\texttt{underflow-read-ptr-heap} \\
\texttt{underflow-read-ptr-data} \\
\texttt{underflow-read-ptr-rodata} \\
\midrule
\end{tabular}\end{footnotesize}

Both buffer \texttt{overflow} and \texttt{underflow} reads have been tested.
We consider two potential ways to access data out-of-boundary:
One is to access with a valid buffer pointer but an out-of-boundary offset (\texttt{index})
and the other one is to directly make a previously valid buffer pointer point
to an out-of-boundary location (\texttt{ptr}).
For defenses using memory tags~\cite{Qin2005} or fat pointers~\cite{Devietti2008, Nagarakatte2009, Kwon2013},
these two ways may result in exceptions being triggered at different locations depending on whether the pointer~\cite{Cowan2003a}
or the buffer memory~\cite{Qin2005} is tagged.
Since the defense mechanisms on stack, heap, global data and read-only data sections might be different,
they are tested separately.
These tests are the most general cases for out-of-boundary reads
that read only the neighbors of the buffer (overflow by one) without going too far.

\subsubsection{General write out-of-boundary}
12 test cases in total.\\
\begin{footnotesize}\begin{tabular}{|l|}
\midrule
\texttt{overflow-write-index-stack} \\
\texttt{overflow-write-index-heap} \\
\texttt{overflow-write-index-data} \\
\texttt{overflow-write-ptr-stack} \\
\texttt{overflow-write-ptr-heap} \\
\texttt{overflow-write-ptr-data} \\
\texttt{underflow-write-index-stack} \\
\texttt{underflow-write-index-heap} \\
\texttt{underflow-write-index-data} \\
\texttt{underflow-write-ptr-stack} \\
\texttt{underflow-write-ptr-heap} \\
\texttt{underflow-write-ptr-data} \\
\midrule
\end{tabular}\end{footnotesize}

Similar to out-of-boundary read tests, these tests try to write out-of-boundary.
Both buffer overflow and underflow writes have been tested.
Both index and pointer ways of accesses have been considered.
Buffers on stack, heap and global data sections have been tested separately.

\subsubsection{Access cross the object boundary}
20 test cases in total.\\
\begin{footnotesize}\begin{tabular}{|l|}
\midrule
\texttt{read-cross-object-index-stack} \\
\texttt{read-cross-object-index-heap} \\
\texttt{read-cross-object-index-data} \\
\texttt{read-cross-object-index-rodata} \\
\texttt{read-cross-object-ptr-stack} \\
\texttt{read-cross-object-ptr-heap} \\
\texttt{read-cross-object-ptr-data} \\
\texttt{read-cross-object-ptr-rodata} \\
\texttt{overflow-write-cross-object-index-stack} \\
\texttt{overflow-write-cross-object-index-heap} \\
\texttt{overflow-write-cross-object-index-data} \\
\texttt{overflow-write-cross-object-ptr-stack} \\
\texttt{overflow-write-cross-object-ptr-heap} \\
\texttt{overflow-write-cross-object-ptr-data} \\
\texttt{underflow-write-cross-object-index-stack} \\
\texttt{underflow-write-cross-object-index-heap} \\
\texttt{underflow-write-cross-object-index-data} \\
\texttt{underflow-write-cross-object-ptr-stack} \\
\texttt{underflow-write-cross-object-ptr-heap} \\
\texttt{underflow-write-cross-object-ptr-data} \\
\midrule
\end{tabular}\end{footnotesize}

All tests are special cases of the \emph{general read out-of-boundary} and
the \emph{general write out-of-boundary} cases.
Some defenses may choose to put invalid data between objects as a tripwire or
enforce the boundary check at the granularity of objects~\cite{Devietti2008}.
In these cases, the general overflow vulnerabilities exist but
the more severe form of them, accessing crossing the object boundary, is disabled.
To check these types of vulnerabilities,
the test suite tests both read and write accesses,
both index and pointer ways of accessing,
and objects on stack, heap, global data and read-only data.

\subsubsection{Access cross the stack frame}
Four test cases in total.\\
\begin{footnotesize}\begin{tabular}{|l|}
\midrule
\texttt{read-cross-frame-index} \\
\texttt{read-cross-frame-ptr} \\
\texttt{write-cross-frame-index} \\
\texttt{write-cross-frame-ptr} \\
\midrule
\end{tabular}\end{footnotesize}

All tests are special cases of the \emph{general read out-of-boundary} and
the \emph{general write out-of-boundary} cases.
To thwart overflow attacks against variables over the frame boundary,
some defenses choose to put padding bytes between the frames~\cite{Bhatkar2005}
while some other defenses use fat-pointers to enforce data integrity at the frame granularity~\cite{Das2019}.
Data compartmentalization at the frame granularity is also effective~\cite{Nyman2019}.
The test suite tests both read and write accesses across the frame boundary using indices or pointers.

\subsubsection{Access cross pages}
Four test cases in total.\\
\begin{footnotesize}\begin{tabular}{|l|}
\midrule
\texttt{read-cross-page-index} \\
\texttt{read-cross-page-ptr} \\
\texttt{write-cross-page-index} \\
\texttt{write-cross-page-ptr} \\
\midrule
\end{tabular}\end{footnotesize}

All tests are special cases of the \emph{access cross the stack frame} cases.
Page table attributes have long been used to enforce access permissions.
System software or memory allocation library may use the page accessing property to partially thwart overflow attacks.
Since pages are normally larger than stack frames,
we consider these as special cases of \emph{access cross the stack frame}.
The test suite tests both read and write access cross the page boundary using indices or pointers.

\subsubsection{Access cross sections}
36 test cases in total.\\
\begin{footnotesize}\begin{tabular}{|l|}
\midrule
\texttt{read-cross-section-stack-to-heap-index} \\
\texttt{read-cross-section-stack-to-data-index} \\
\texttt{read-cross-section-stack-to-rodata-index} \\
\texttt{read-cross-section-heap-to-stack-index} \\
\texttt{read-cross-section-heap-to-data-index} \\
\texttt{read-cross-section-heap-to-rodata-index} \\
\texttt{read-cross-section-data-to-stack-index} \\
\texttt{read-cross-section-data-to-heap-index} \\
\texttt{read-cross-section-data-to-rodata-index} \\
\texttt{read-cross-section-rodata-to-stack-index} \\
\texttt{read-cross-section-rodata-to-heap-index} \\
\texttt{read-cross-section-rodata-to-data-index} \\
\texttt{read-cross-section-stack-to-heap-ptr} \\
\texttt{read-cross-section-stack-to-data-ptr} \\
\texttt{read-cross-section-stack-to-rodata-ptr} \\
\texttt{read-cross-section-heap-to-stack-ptr} \\
\texttt{read-cross-section-heap-to-data-ptr} \\
\texttt{read-cross-section-heap-to-rodata-ptr} \\
\texttt{read-cross-section-data-to-stack-ptr} \\
\texttt{read-cross-section-data-to-heap-ptr} \\
\texttt{read-cross-section-data-to-rodata-ptr} \\
\texttt{read-cross-section-rodata-to-stack-ptr} \\
\texttt{read-cross-section-rodata-to-heap-ptr} \\
\texttt{read-cross-section-rodata-to-data-ptr} \\
\texttt{write-cross-section-stack-to-heap-index} \\
\texttt{write-cross-section-stack-to-data-index} \\
\texttt{write-cross-section-heap-to-stack-index} \\
\texttt{write-cross-section-heap-to-data-index} \\
\texttt{write-cross-section-data-to-stack-index} \\
\texttt{write-cross-section-data-to-heap-index} \\
\texttt{write-cross-section-stack-to-heap-ptr} \\
\texttt{write-cross-section-stack-to-data-ptr} \\
\texttt{write-cross-section-heap-to-stack-ptr} \\
\texttt{write-cross-section-heap-to-data-ptr} \\
\texttt{write-cross-section-data-to-stack-ptr} \\
\texttt{write-cross-section-data-to-heap-ptr} \\
\midrule
\end{tabular}\end{footnotesize}

All tests are special cases of the \emph{general read out-of-boundary} and
the \emph{general write out-of-boundary} cases.
Attackers can manipulate the stack data using a heap pointer~\cite{Conti2015} and probably vice versa.
To test these special attack variants,
the test suite checks all possible scenarios for accessing cross section boundaries.

\subsubsection{Spray attack}
Six test cases in total.\\
\begin{footnotesize}\begin{tabular}{|l|}
\midrule
\texttt{spray-cross-object-stack} \\
\texttt{spray-cross-object-heap} \\
\texttt{spray-cross-object-data} \\
\texttt{spray-cross-frame} \\
\texttt{spray-cross-page-stack} \\
\texttt{spray-cross-page-heap} \\
\midrule
\end{tabular}\end{footnotesize}

All tests are special cases of the \emph{general write out-of-boundary} cases.
In the access tests, out-of-boundary data are directly accessed using indices or pointers and
leave all the data in between untouched.
While in spray tests, all the data between the buffer to the out-of-boundary target are rewritten.
Such behavior is more likely to be detected by memory padding~\cite{Bhatkar2005, Devietti2008} and tagging~\cite{Venkataramani2007}
than a direct out-of-boundary access.
The test suite currently tests six spray attacks:
spray cross the object boundary on stack, heap and global data,
spray cross stack frames,
and spray cross pages on stack and heap.

\subsection{Temporal Safety}

Temporal safety refers to the property that data accessed in memory happens only within the lifetime of the data.
Any access before (uninitialized) or after (use-after-free, UAF) the lifetime is considered insecure.
These types of vulnerabilities potentially lead to corrupted values in memory~\cite{Chen2005},
return-to-libc~\cite{Pincus2004}, and arbitrary memory read and write eventually.

Since the target of this test suite is to measure the safety supported by the processor hardware
as described in Section~\ref{subsec:scope},
its test cases concentrate on whether a general out-of-lifetime access can happen and
consider attacks specialized to certain memory allocation algorithms~\cite{Xie2016} out of the scope.

\subsubsection{Access after free on heap}
Three test cases in total.\\
\begin{footnotesize}\begin{tabular}{|l|}
\midrule
\texttt{read-after-free-org-heap} \\
\texttt{read-after-free-alias-heap} \\
\texttt{write-after-free-heap} \\
\midrule
\end{tabular}\end{footnotesize}

After an object is released, all pointers pointing to this object become dangling pointers and should not be used afterward.
However, attackers may utilize these dangling pointers to read the previously released object, leading to information leakage.
Even worse, attackers may write to the just released memory which potentially causes data corruption.
To thwart these threats, defenses can deliberately nullify all dangling pointers~\cite{Lee2015, Younan2015}
or checking whether a pointer is dangling at the dereference point~\cite{Nagarakatte2012, Nagarakatte2014}.
The three tests here exam the possibility of reading/writing a heap object after it is released.
For reading the released object, both the original pointer or a copy of it (alias) have been tested
as a way to measure the completeness of pointer nullification.

\subsubsection{Reclaim on heap}
Only one test case.\\
\begin{footnotesize}\begin{tabular}{|l|}
\midrule
\texttt{reallocate-heap} \\
\midrule
\end{tabular}\end{footnotesize}

Once a dangling pointer falls into the control of attackers,
it can be used to launch targeted data corruption
if the released memory is reallocated to the same type of objects.
Some secure memory allocators try to prevent objects from reclaiming the previously released memory~\cite{Novark2010,Akritidis2010}
but attackers may force the memory allocator to do so~\cite{Lee2015}.
To measure this vulnerability,
this test tries to force the memory allocator to place an object with the same type at the same location
where it is previously released.

\subsubsection{Access after reclaim on heap}
Three test cases in total.\\
\begin{footnotesize}\begin{tabular}{|l|}
\midrule
\texttt{access-after-reclaim-heap} \\
\texttt{write-before-reclaim-heap} \\
\texttt{write-after-reclaim-heap} \\
\midrule
\end{tabular}\end{footnotesize}

All tests depend on the \emph{reclaim on heap} test.
Assuming attackers have successfully forced an object to reclaim the same memory space previously released,
they would then try to manipulate the object using a dangling pointer.
\texttt{\seqsplit{access-after-reclaim-heap}} tests whether
an uninitiated variable belonging to the newly allocated object retains its previous value.
In case the memory is safely cleaned when it is released,
\texttt{write-before-reclaim-heap} tries to maliciously initiate this uninitiated variable just before the memory is reclaimed.
Finally, \texttt{write-after-reclaim-heap} tests
whether a dangling pointer can be used to corrupt an object after it reclaims the same memory space.

\subsubsection{Access after free on stack}
Two test cases in total.\\
\begin{footnotesize}\begin{tabular}{|l|}
\midrule
\texttt{read-after-free-alias-stack} \\
\texttt{write-after-free-stack} \\
\midrule
\end{tabular}\end{footnotesize}

UAF attacks happen on stack as well.
When a function returns,
the stack frame is released for the next function call.
If a dangling pointer pointing to the released stack frame is mistakenly leaked
by a function argument, a global variable or even the return value,
it allows attackers to manipulate the frame when it is reclaimed~\cite{Younan2015}.
Similar to the tests for UAF on heap,
the tests start with reading a released stack variable to see if it retains its value
and then write it to see whether the released frame space is protected.

\subsubsection{Reclaim on stack}
Only one test case.\\
\begin{footnotesize}\begin{tabular}{|l|}
\midrule
\texttt{reallocate-stack} \\
\midrule
\end{tabular}\end{footnotesize}

The memory structure of the stack frame for the same function is normally the same every time the function is called.
This potentially allows attackers to precisely manipulate the variables on the stack using a dangling pointer.
Fine-grained runtime stack layout randomization~\cite{Aga2019} may stop such attacks by
locating stack variables at different positions whenever a function is called again.
This test verifies whether the stack layout is changed when a function is called again.

\subsubsection{Access after reclaim on stack}
Three test cases in total.\\
\begin{footnotesize}\begin{tabular}{|l|}
\midrule
\texttt{access-after-reclaim-stack} \\
\texttt{write-before-reclaim-stack} \\
\texttt{write-after-reclaim-stack} \\
\midrule
\end{tabular}\end{footnotesize}

All tests are dependent on the \emph{reclaim on stack} test.
Assuming the stack layout remains the same for the same function,
\texttt{\seqsplit{access-after-reclaim-stack}} tests whether an uninitiated variable keeps its value from the previous function call,
\texttt{\seqsplit{write-before-reclaim-stack}} tries to maliciously initiate the uninitiated variable before the next function call.
Finally, \texttt{write-after-reclaim-stack} tests
whether a dangling pointer can be used to alter a stack variable when the stack frame is reclaimed by a new function call.

\subsection{Access Control}

We consider all defense techniques that exert restrictions on accesses as \emph{access control} related defenses.
ASLR~\cite{Forrest1997} normally acts as the first level of defense against code-reuse attacks.
It prohibits attackers from collecting the addresses of the required gadgets by static binary analysis.
DEP is another widely adopted defense that disallows codes in writable pages from being executed.
To thwart the dynamic code-reuse attacks which collect gadgets at runtime~\cite{Snow2013},
both code randomization~\cite{Sinha2017} and read exclusive execution (R$\oplus$X)~\cite{Pomonis2019}
have been proposed to forbid attacker from reading executable pages.

\subsubsection{Access control}
Three test cases in total.\\
\begin{footnotesize}\begin{tabular}{|l|}
\midrule
\texttt{check-ASLR} \\
\texttt{read-func} \\
\texttt{read-GOT} \\
\midrule
\end{tabular}\end{footnotesize}

\texttt{check-ASLR} checks whether the address space is randomized,
\texttt{read-func} checks whether the body of a function (in a code page) can be read,
and finally \texttt{read-GOT} checks whether an entry in the global offset table (GOT) can be read.

\subsection{Pointer Integrity}

We consider all attacks that directly cause a malicious alteration of control flow as control-flow related attacks
and treat all defense techniques against such attacks as control flow related defenses.
Most defenses fall into two categories: \emph{pointer integrity} and \emph{control-flow integrity}.
The former prevents pointers (mostly code related) from being maliciously modified~\cite{Kuznetsov2014}
while the latter tries to stop those altered pointers from affecting the control flow~\cite{Abadi2009}.

Since most control-flow attacks rely on the alteration of certain key pointers,
the tests for pointer integrity are considered as prerequisites for the tests for control-flow integrity.
This set of tests aim to cover the alteration of different types of key pointers.
Note that the alteration of return addresses is closely related to ROP attacks and is thus tested there.

\subsubsection{Function pointer}
Two test cases in total.\\
\begin{footnotesize}\begin{tabular}{|l|}
\midrule
\texttt{func-pointer-assign} \\
\texttt{func-pointer-arithmetic} \\
\midrule
\end{tabular}\end{footnotesize}

Function pointers are one of the most frequently attacked targets.
\texttt{\seqsplit{func-pointer-assign}} checks whether a function pointer can be altered with embedded assembly
(therefore bypassing the analysis and potential protections from compilers
through pointer authentication~\cite{Mashtizadeh2015} or tagging~\cite{Kuznetsov2014, Song2016}).
It is very rare that arithmetic operations occur on function pointers~\cite{Chen2019}.
Function pointers are normally cloneable but not mutable.
By applying unnecessary arithmetic operations on a function pointer,
\texttt{func-pointer-arithmetic} tests whether such behavior raises any exceptions.

\subsubsection{VTable pointer}
Two test cases in total.\\
\begin{footnotesize}\begin{tabular}{|l|}
\midrule
\texttt{read-vtable-pointer} \\
\texttt{write-vtable-pointer} \\
\midrule
\end{tabular}\end{footnotesize}

Virtual table (VTable) pointers are the major targets in COOP attacks~\cite{Schuster2015}.
Defenses like CPI can protect VTable pointers just like function pointers~\cite{Kuznetsov2014}
and prevent them from being modified.
This protection is tested by \texttt{write-vtable-pointer}.
Since VTable pointers should not be explicitly accessed by any source-level program,
and COOP attacks usually need to read and reuse VTable pointers,
\texttt{read-vtable-pointer} checks whether deliberately reading a VTable pointer would trigger any exceptions.

\subsubsection{Global offset table}
Only one test case.\\
\begin{footnotesize}\begin{tabular}{|l|}
\midrule
\texttt{modify-GOT} \\
\midrule
\end{tabular}\end{footnotesize}

GOT is used by dynamically linked programs to search for
(function and data) symbols defined in shared libraries at runtime.
Since the content of the GOT is updated at runtime,
GOT might be stored on a writable page.
This potentially allows attackers to hijack library functions by altering the corresponding GOT entries~\cite{c0ntex}.
The provided test \texttt{modify-GOT} checks whether an entry in the GOT can be hijacked to an attacker controlled function.

\subsection{Control-Flow Integrity}

Control flow refers to the indirect jumps caused by function returns, function calls and other compiler added jumps.
By altering the return address stored on the stack,
attackers can redirect the control flow to code chosen or injected by them,
which is the basis for ROP attacks~\cite{Shacham2007}.
Similarly, maliciously affecting the pointers used in these indirect calls and jumps~\cite{Bletsch2011, Schuster2015}
can hijack the control flow as well.
Currently, the test suite has 41 test cases to measure the backward control-flow integrity:

\subsubsection{Return to injected code}
Four test cases in toal.\\
\begin{footnotesize}\begin{tabular}{|l|}
\midrule
\texttt{return-to-instruction-in-stack} \\
\texttt{return-to-instruction-in-heap} \\
\texttt{return-to-instruction-in-data} \\
\texttt{return-to-instruction-in-rodata} \\
\midrule
\end{tabular}\end{footnotesize}

All tests are special cases for \emph{return-to-non-call-site}.
Although attacks relying on direct code injection are largely thwarted by DEP,
the same defense may not be available on legacy systems and embedded systems.
As a result, this test suite still checks whether a function can be hijacked and returned to an injected code.
Note that injecting code using constant texts (read-only data) remains viable on most systems
unless some forms of R$\oplus$X is enforced~\cite{Pomonis2019}.

\subsubsection{Return to existing code}
Six test cases in total.\\
\begin{footnotesize}\begin{tabular}{|l|}
\midrule
\texttt{return-to-wrong-call-site-within-static-analysis} \\
\texttt{return-to-wrong-call-site} \\
\texttt{return-to-non-call-site} \\
\texttt{return-to-func} \\
\texttt{return-to-libc} \\
\texttt{return-without-call} \\
\midrule
\end{tabular}\end{footnotesize}

All tests are relaxed cases of \texttt{\seqsplit{return-to-wrong-call-site-within-static-analysis}}.
\texttt{\seqsplit{return-to-non-call-site}} represents the common ROP attacks that use arbitrary code snippets as gadgets.
To avoid the detection of coarse-grained CFI checks,
attackers may be forced to use call preceded gadgets (\texttt{\seqsplit{return-to-wrong-call-site}})~\cite{Carlini2014},
long and benign code segment (\texttt{return-to-func}~\cite{Goektas2014} and \texttt{return-to-libc}~\cite{Pincus2004}),
or even previously returned locations (replay attack, \texttt{\seqsplit{return-to-wrong-call-site-within-static-analysis}}~\cite{Roessler2018}).
When multiple gadgets are chained together,
the call-return pair becomes unbalanced, which should be easy to detect by a shadow stack~\cite{Dang2015, Burow2019}.
This type of unbalance is tested by \texttt{return-without-call}.

\subsubsection{Call to injected code}
Four test cases in total.\\
\begin{footnotesize}\begin{tabular}{|l|}
\midrule
\texttt{call-instruction-in-stack} \\
\texttt{call-instruction-in-heap} \\
\texttt{call-instruction-in-data} \\
\texttt{call-instruction-in-rodata} \\
\midrule
\end{tabular}\end{footnotesize}

All tests are special cases of \texttt{call-mid-func}.
Similar to the tests for backward control-flow integrity,
these tests check whether a function call can be hijacked to a code injected by attackers.

\subsubsection{Call to existing code}
Three test cases in total.\\
\begin{footnotesize}\begin{tabular}{|l|}
\midrule
\texttt{call-wrong-func-within-static-analysis} \\
\texttt{call-wrong-func} \\
\texttt{call-mid-func} \\
\midrule
\end{tabular}\end{footnotesize}

All tests are special cases of \texttt{\seqsplit{call-wrong-func-within-static-analysis}}.
\texttt{call-mid-func} is the most general case as the function pointer is hijacked to an arbitrary location.
Such attacks can be easily detected by most CFI checks~\cite{Abadi2009, Burow2017}
as the arbitrary location is rarely a function entry point.
To avoid detection by coarse-grained CFI checks~\cite{Davi2014},
\texttt{call-wrong-func} hijacks a function pointer to another function
while the function chosen by \texttt{call-wrong-func-within-static-analysis}
also falls in the valid set of targets collected by static control-flow analyses
(thus is detectable only to path-sensitive CFI defenses~\cite{Ding2017, Zhang2018}).

\subsubsection{Call with wrong arguments}
Nine test cases in total.\\
\begin{footnotesize}\begin{tabular}{|l|}
\midrule
\texttt{call-wrong-num-arg-func} \\
\texttt{call-wrong-type-arg-int2double-func} \\
\texttt{call-wrong-type-arg-op2doublep-func} \\
\texttt{call-wrong-type-arg-fp2dp-func} \\
\texttt{call-wrong-type-arg-dp2fp-func-stack} \\
\texttt{call-wrong-type-arg-dp2fp-func-heap} \\
\texttt{call-wrong-type-arg-dp2fp-func-data} \\
\texttt{call-wrong-type-arg-dp2fp-func-rodata} \\
\midrule
\end{tabular}\end{footnotesize}

All the \texttt{\seqsplit{call-wrong-type-arg-dp2fp-func-xxxx}} test cases
depend on the \texttt{\seqsplit{call-instruction-in-xxxx}}.
Some research claims that even the fine-grained CFI defenses can be circumvented
by corrupting only the arguments of functions~\cite{Evans2015}
while the types of arguments can be used to enhance the accuracy in detecting COOP attacks~\cite{Veen2016}.
To cover these attack and defense variants,
\texttt{call-wrong-num-arg-func}
check whether a function can be called with a mismatched number of arguments.
Extra tests are provided for function calls with mismatched types of arguments,
such as providing an integer for a double argument (\texttt{int2double}),
providing an object pointer for a double or an integer pointer (\texttt{op2doublep} and \texttt{op2intp}),
and providing a function pointer to a data pointer (\texttt{fp2dp}).
The last four tests check whether injected code can be called by placing a data pointer in the place for a function pointer (\texttt{dp2fp}).

\subsubsection{VTable injection}
Three test cases in total.\\
\begin{footnotesize}\begin{tabular}{|l|}
\midrule
\texttt{call-wrong-func-vtable-stack} \\
\texttt{call-wrong-func-vtable-heap} \\
\texttt{call-wrong-func-vtable-data} \\
\midrule
\end{tabular}\end{footnotesize}

Modern compilers put VTables on read-only pages to prevent them from being altered.
However, attackers can forge fake VTables in other writable spaces
and using them to dislodge the real ones~\cite{Schuster2015}.
These three tests check whether VTables can be replaced with a fake one created
on stack, heap and global data.

\subsubsection{Replace VTables with existing tables}
Seven test cases in total.\\
\begin{footnotesize}\begin{tabular}{|l|}
\midrule
\texttt{call-wrong-func-vtable-parent} \\
\texttt{call-wrong-func-vtable-child} \\
\texttt{call-wrong-func-vtable-sibling} \\
\texttt{call-wrong-func-vtable} \\
\texttt{call-wrong-func-vtable-released} \\
\texttt{call-wrong-func-vtable-offset} \\
\texttt{call-wrong-num-arg-vtable} \\
\midrule
\end{tabular}\end{footnotesize}

Assuming VTables cannot be forged,
an attacker would try replacing them with existing ones inside the memory.
To thwart this type of COOP attacks,
class hierarchy analysis~\cite{Jang2014} tries to limit the tables that are available to attackers.
\texttt{call-wrong-func-vtable} checks the most general case where a VTable is replaced by an arbitrary table existing in memory.
The test suite then checks whether the VTables from a \texttt{parent} class,
a \texttt{child} class or a \texttt{sibling} class can be used for replacement.
\texttt{call-wrong-func-vtable-released} tests whether a VTable pointer of a released object
can be reused in a live object~\cite{Sarbinowski2016}.
\texttt{call-wrong-func-vtable-offset} checks
whether a VTable pointer can be added with a small offset.
Finally, \texttt{call-wrong-num-arg-vtable} checks whether the virtual function
can be replaced with another one with different number arguments;
therefore, allowing attackers to manipluate the arguments.

\subsubsection{Indirect jump}
Five test cases in total.\\
\begin{footnotesize}\begin{tabular}{|l|}
\midrule
\texttt{jump-instruction-in-stack} \\
\texttt{jump-instruction-in-heap} \\
\texttt{jump-instruction-in-data} \\
\texttt{jump-instruction-in-rodata} \\
\texttt{jump-mid-func} \\
\midrule
\end{tabular}\end{footnotesize}

\texttt{jump-mid-func} checks whether an arbitrary code snippet existing in memory can be used by a general JOP attack.
All \texttt{\seqsplit{jump-instruction-in-xxxx}} test cases are special cases of \texttt{jump-mid-func}.
Similar to the tests for ROP and COOP attacks,
these four tests check whether a JOP attack~\cite{Bletsch2011} can hijack the jump target to an injected code.

\section{Challenges in Implementation}\label{sec:challenge}

In this section, we describe the challenges encountered during the implementation 
and our solutions to resolve them. 

\subsection{Obtain Parameters at Runtime}\label{subsec:runtime-var}

Some test parameters are only reliably available at runtime.
One such example is the offset of the return address on the stack from the stack pointer,
as mentioned in Section~\ref{subsec:con-tc}.
The value of offset depends on the ABI (application binary interface) definition of the target architecture,
whether the frame pointer is pushed on the stack,
and whether code instrument affects the stack,
such as initializing the canary for stack smashing protection~\cite{Cowan2003}.
The value of this offset is therefore hardly fixed but
all ROP related test cases rely on this parameter.

In this test suite,
the value of offset is obtained from a blind testing using \texttt{return-to-wrong-call-site},
which is the common prerequisite for all ROP tests,
and fed to all depended tests as an input argument.
As shown in Listing~\ref{list:test-case} and described in Section~\ref{subsec:con-tc},
the test gets an \texttt{offset} from input and checks whether the return address
can be hijacked to a seemly valid return site \texttt{main\_mid}.
Since this \texttt{offset} is unknown,
the test suite tries to get the correct value by launching
the same test multiple times with different values of \texttt{offset}.
Such behavior is defined in the configuration file through an extensions illustrated in Listing~\ref{list:configure-runtime-var}.

\begin{figure}[bt]
\centering{
\begin{minipage}{0.43\textwidth}
\begin{lstlisting}[language=json, caption=\texttt{Feed runtime parameter between test cases}, label=list:configure-runtime-var]
{
  ... // omit some code

  "return-to-wrong-call-site": {
    ... // omit some code
    "arguments": ["-roffset"],
    "offset": [0, 8, 1],
    "set-var": "stack-offset"
  },

  "return-to-non-call-site": {
    "require": [ [ "return-to-wrong-call-site" ] ],
    "arguments": ["-vstack-offset"]
  },
  
  ... // omit some code
}
\end{lstlisting}
\end{minipage}
}
\end{figure}

The list recorded in property \texttt{arguments}  (line 6  in Listing~\ref{list:configure-runtime-var})
defines the input arguments for a test.
Arguments are simply attached to the test case except for the special ones starting with ``\texttt{-}''.
These special arguments always take the form of ``\texttt{-}$t$\textsc{name}'',
where $t$ denotes the type and \textsc{name} identifies the property defining this argument.
Currently, we support three types: range (`\texttt{r}'), list (`\texttt{l}) and variable (`\texttt{v}'). 
When an argument takes the form of ``\texttt{-}$r$\textsc{name}'', it actually defines a range (``\texttt{-r}'')
defined in a property named ``\textsc{name}''.
In the case of \texttt{return-to-wrong-call-site}, 
a range of [0:7] (\texttt{[0,8,1]} in python) is defined by property \texttt{offset}(line 8  in Listing~\ref{list:configure-runtime-var})
and used as an argument, which means the test would be launched eight times for each value in the range.
The final property \texttt{set-var} denotes that
the variable \texttt{stack-offset} would be assigned with the \texttt{offset} resulting a success test.
For example, if test \texttt{return-to-wrong-call-site} succeeds with argument 3,
\texttt{stack-offset} is then set to 3 and
all other test cases using this variable would get an argument of 3.
\texttt{return-to-non-call-site} is one of these cases (line 11).
Its input argument is therefore defined to use the value of \texttt{stack-offset}
(line 13 in Listing~\ref{list:configure-runtime-var}).
Since \texttt{\seqsplit{return-to-wrong-call-site}} is listed as a prerequisite, 
the value of \texttt{stack-offset} must have been assigned when \texttt{\seqsplit{return-to-non-call-site}} is tested.
In other words, the test scheduler and the JSON configuration
ensure that \texttt{stack-offset} is initiated before testing all the test cases depending on it.

\subsection{Fight Against Compiler Optimization}\label{subsec:cpp-opt}

Since test cases usually contain deliberately malformed code and
the behavior of the embedded assembly is commonly unexpected by compilers,
we often find that the attacking behavior of a test case is silently disarmed by compiler optimizations.
To avoid this, the test suite utilizes certain code patterns and compiler directives
to disable compiler optimization at precise locations rather than
forcefully applying compiler flags to disabling certain optimization to all code.
Here we describe some of the techniques utilized in this test suite.

When a function is simple enough or seemly dead, with or without an embedded assembly,
compiler would trying to eliminate the function call by function inlining, dead code elimination, constant propagation,
or even return value prediction.
To prevent these optimizations,
we use compiler attributes to disable the inlining of key functions. 
Some key variables are declared \texttt{volatile} to stop dead code elimination and return value prediction.
In some extreme cases, key values are fed to the test case through input arguments to avoid constant propagation.

Compiler may choose to reorder instructions which breaks the assumptions in certain test cases.
One of such examples is the \texttt{read-GOT} test shown in Listing~\ref{list:read-got}.
Function \texttt{get\_got\_func()} on  line 3 
tries to obtain the address of the GOT entry for \texttt{rand()}.
A reliable way to get this address is to call \texttt{rand()} immediately after \texttt{get\_got\_func()}.
The address of the corresponding PLT (procedure linkage table) entry is then saved as the return address for \texttt{get\_got\_func()}.
To our surprise, some Arm variants of GCC swap the calling of \texttt{rand()}
and the following dereference of the obtained GOT address (\texttt{got}),
which usually leads to a segment error,
resulting in an incorrect detection of a non-existent defense.
To avoid such unexpected behavior, which might lead to an incorrect detection of a non-existent defense,
we add a compiler barrier on line 5 to prevent instruction reorder on this particular location.

\begin{figure}[bt]
\centering{
\begin{minipage}{0.43\textwidth}
\begin{lstlisting}[language=c++, caption=\texttt{Read GOT entries}, label=list:read-got]
// part of main() in read-GOT.cpp
void *got = NULL;
get_got_func(&got, offset);
rand();
COMPILER_BARRIER;
return *(uintptr_t *)(got);

// part of assembly.hpp
#define COMPILER_BARRIER \
  asm volatile("" : : : "memory")
\end{lstlisting}
\end{minipage}
}
\end{figure}

Another interesting case is related to the hijacking of VTable pointers as illustrated in Listing~\ref{list:fake-vtable}.
The genuine VTable of the victim object \texttt{orig} is replaced with the VTable from an irrelevant object \texttt{fake}
on line 11.
Supposing this replacement is successful,
the following \texttt{orig->virtual\_func()} should call the wrong function defined in \texttt{Fake}
which exits with a code 0.
However, the compiler seemly believes \texttt{orig->virtual\_func()} should always call the genuine virtual function
and replaces the VTable based indirect call with a direct call by calculating an offset at compile time.
The result is a failed test case with an arguably successful attack.
To avoid such compile time (possibly link time as well) optimization,
we hide the definition of the victim classes to a shared library which is visible only at runtime.

\begin{figure}[bt]
\centering{
\begin{minipage}{0.43\textwidth}
\begin{lstlisting}[language=c++, caption=\texttt{Fake VTable}, label=list:fake-vtable]
... // omit some code
class Fake {
public:
  virtual void virtual_func() { exit(0); }
  ... // omit some code
};

int main() {
  Base *orig = new Base();
  Fake *fake = new Fake();
  write_vtable_pointer(orig, *((pvtable_t *)fake));
  orig->virtual_func();
  return 4;
}
\end{lstlisting}
\end{minipage}
}
\end{figure}

\subsection{Interpret Test Results}

It is actually a big challenge to interpret the results of test cases.
As indicated by the examples shown in Section~\ref{subsec:cpp-opt},
unexpected compiler optimizations may fail a test case and indicate the existence of an actually non-existent defense.
In fact, the reverse could happen as well,
such as that a defense may replace a dangling pointer with a random number pointing to a still readable memory location.
Although some coding techniques have been applied to avoid these errors, removing all of them for all compilers is difficult.
When a test indeed fails due to a defense,
it is also hard to tell who implements this defense, the compiler, the runtime library, the kernel, or the hardware.
Considering a ROP attack as an example,
it might be defeated by software implemented canary or a shadow stack,
or hardware supported pointer authentication or memory tagging.
In this test suite, two mechanisms are utilized to reduce ambiguity and potentially warn about misinterpretation:

(1) Different exit codes are used for all exiting points of the program,
such as the code 4 used on line 13 of Listing~\ref{list:fake-vtable}.
Code 0 is returned only when the test is considered successful.
Exit codes for known defenses are listed in the configuration file,
such as the code 16 in Listing~\ref{list:exit-code}.
During the test, the scheduler silents the output for all tests returning 0 or known codes
while highlighting the results for tests returning unexpected code
as a way to indicate the need for further investigation.

\begin{figure}[bt]
\centering{
\begin{minipage}{0.43\textwidth}
\begin{lstlisting}[language=json, caption=\texttt{Record expected exit codes}, label=list:exit-code]
{
  ... // omit some code

  "code-injection-stack": {
    ... // omit some code
    "results": { "16": "DEP protection" }
  },

  ... // omit some code
}
\end{lstlisting}
\end{minipage}
}
\end{figure}

(2) When a test ends with an exception,
the test suite tries to catch it with extra checks.
An example (case \texttt{\seqsplit{return-to-instruction-in-stack}}) is demonstrated in Listing~\ref{list:exception},
which checks whether the return address can be hijacked to a code injected on stack by a ROP attack.
Similar to \texttt{\seqsplit{return-to-wrong-call-site}} (Listing~\ref{list:test-case}),
the actual attack is launched by \texttt{helper(m)} on line 6.
On most platforms, such attack is easily thwarted by DEP.
A segment error is thrown immediately when the address of the injected code (\texttt{m}) is assigned to PC.
This behavior is caught by the outer pair of \texttt{catch\_exception()} functions,
which specifically watches the \texttt{SEGV\_ACCERR} error on address \texttt{m}.
The test then exits with code 16 to indicate DEP (with test termination by exception).
In the rare condition when stack is executable,
the test would succeed with executing the code on stack.
To improve the platform independence and avoid recovery from a corrupted call stack,
the test injects a divide-by-zero instruction sequence with the division instruction on \texttt{m+4}. 
The division would raise an exception of type \texttt{SIGFPE},
which is caught by the inner pair of \texttt{catch\_exception()}.
The exit code is then set to 0.
All codes other than 0 and 16 would need further investigation.

\begin{figure}[bt]
\centering{
\begin{minipage}{0.43\textwidth}
\begin{lstlisting}[language=c++, caption=\texttt{Code injection (stack)}, label=list:exception]
// part of main() in
// return-to-instruction-in-stack.cpp
unsigned char m[] = DIVIDE_BY_0_CODE;
begin_catch_exception(m, SEGV_ACCERR, 16, SIGSEGV);
 begin_catch_exception(m+4, 0, 0, SIGFPE);
  int rv = helper(m); // hijacted to m by ROP
 end_catch_exception();
end_catch_exception();
exit(rv);

// definition of begin_catch_exception
// in assembly.hpp
extern void begin_catch_exception(
  void *addr, int si_code,
  int exit_code, int si_signo);
\end{lstlisting}
\end{minipage}
}
\end{figure}

\section{Test Results}\label{sec:result}


\subsection{Testing Platforms}

\begin{table*}[bt]
  \footnotesize
  \centering
  \caption{Platform parameters}\label{tab:platform}
  \begin{tabular}{lccccccc}
    \toprule
    Platform           & Architecture & Processor        & Operating System & Kernel  & Compiler   & C Library\\
    \midrule
    Intel i7-3770      & x86-64         & i7-3770        & Ubuntu 16.04     & 4.15.0  & g++ 5.4.0  & GLIBC 2.23\\
    Intel Xeon 8280    & x86-64         & Xeon 8280      & Ubuntu 18.04     & 5.4.0   & g++ 7.5.0  & GLIBC 2.27\\
    Jetson Nano        & ARMv8.0-A      & Arm Cortex-A57 & Ubuntu 18.04     & 4.9.140 & g++ 7.5.0  & GLIBC 2.27\\
    Raspberry Pi 4B    & ARMv8.0-A      & Arm Cortex-A72 & Ubuntu 20.10     & 5.8.0   & g++ 10.2.0 & GLIBC 2.32\\
    HiFive Unleashed   & RV64GC         & SiFive u540    & OpenEmbedded~\cite{SiFive2021meta} & 5.8.2 & g++ 10.2.0 & GLIBC 2.32\\
    Spike              & RV64GC         & Spike 2020-Oct & Buildroot~\cite{Xu-rss-sdk} & 5.8.1 & g++ 10.1.0 & GLIBC 2.32\\
    \bottomrule
  \end{tabular}
\end{table*}

The test suite has been applied on six platforms using three different ISAs,
including Intel x86-64, Arm AArch64 and RISC-V.
Intel x86-64 represents the most used ISA in personal computers (PC) and servers.
Arm AArch64 is currently the most utilized ISA on smartphones.
RISC-V~\cite{Waterman2019} is a newly proposed and promising open architecture
which has been quickly adopted by the opensource community and processor manufactures.

Table~\ref{tab:platform} illustrates the parameters of the six platforms.
For Intel x86-64, a relatively old i7-3770 with Ubuntu 16.04
and a recent Xeon 8280 with Ubuntu 18.04 are chosen as two representative platforms.
We have managed to run the test suite on two Arm single board computers which we have access to.
One is a Jetson Nano board with a Ubuntu 18.04 running on an Arm Cortex-A57 processor
and the other is a Raspberry Pi 4B board with a Ubuntu 20.10 running on an Arm Cortex-A72 processor.
Both processors comply with the ARMv8.0-A ISA but are equipped with different operating systems.
As for RISC-V, we have managed to borrow a HiFive Unleashed board mounted with a SiFive u540 processor~\cite{SiFive2021U540}.
Although HiFive Unleashed was released in 2018, it is still one of the most powerful RISC-V computers ever produced.
The operating system running on the HiFive Unleashed is a standard pre-compilation of the OpenEmbedded Linux~\cite{SiFive2021meta}
recommended by SiFive inc.
To compensate for the lack of a relatively new RISC-V platform,
we have compiled and installed all the necessary libraries in a latest Linux kernel image using buildroot~\cite{Xu-rss-sdk}
and run it on Spike~\cite{Spike},
which is officially the golden reference model and simulator for the latest RISC-V ISA.

\subsection{Comparison Between Platforms}


\begin{table}[bt]
\footnotesize
\begin{center}
  \caption{Test results with the default settings}\label{tab:result-default}
  \begin{tabular}{lcccccc}
    \toprule
    & \rot{\Mlizz} & \rot{\Mstockport} & \rot{\MJetson} & \rot{\MRasp} &  \rot{\MHiFive} & \rot{\MSpike} \\
    \midrule
    Spatial safety (98)     & 98   & 98   & 98   & 98   & 98   & 98 \\
    Temporal safety (13)    & 13   & 13   & 13   & 9    & 9    & 9  \\
    Access control (3)      & 3    & 2    & 2    & 2    & 2    & 3   \\
    Pointer integrity (5)   & 5    & 4    & 4    & 4    & 5    & 5   \\
    CFI (41)                & 28   & 28   & 28   & 28   & 28   & 24  \\
    \midrule
    Total (160)             & 147  & 145  & 145  & 141  & 142  & 139 \\
    \bottomrule
  \end{tabular}
\end{center}
\end{table}

To answer the question \emph{Q1} raised in the introduction,
our first set of tests try to evaluate the security provided by different platforms.
To produce a fair comparison, we use the default setting on all platforms.
The test suite is compiled using the default GNU g++ compiler distributed by the OS
with the compiler flags set as \texttt{-O2 -std=c++11 -Wall}.
Table~\ref{tab:result-default} provides a summary of the test results
while detailed test results are revealed in Appendices~\ref{app:mem-safety}.
Each figure in the table denotes the number of test cases succeed in the specific category.
The success of a test case indicates that one type of vulnerabilities is tested exploitable due to the lack of a memory check.
Consequently, a lower figure indicates better security as less tests succeed.
The name and the total number of cases of each category are provided in the first column of Table~\ref{tab:result-default}.

\subsubsection{Spatial safety}
The results show that \emph{there is currently no defense detected on all platforms
  against the buffer overflow and memory spray attacks tested in this suite.}
All test cases success without exception or error.
This reflects that,
although numerous defense techniques have been proposed and
some of them have been incorporated into compilers in recent years,
no defense is applied by default probably due to performance concerns.

\subsubsection{Temporal safety}
Although most test cases related to temporal safety still pass on all platforms,
all heap-based read-after-free and write-before-reclaim tests fail
on platforms using GLIBC version newer than 2.32.
A detailed look reveals that the memory allocation algorithm comes along with the latest C library
refills the memory space with garbage during both the allocation and the release processes,
which effectively prevent data leakage after free and malicious initialization for uninitiated variables.
However, the new memory allocation algorithm can still be forced to
relocate the same object at the same memory region released previously.
UAF attacks using dangling pointers are still viable.
Meanwhile, stack based UAF attacks are rarely defended.

\subsubsection{Access control}
To our surprise, the tests indicate that ASLR is not deployed on Intel i7-3770 and Spike.
With some further investigation, both platforms do support user-level ASLR
but the default compilers have the position independent executable (PIE) feature disabled by default.
The test case \texttt{check-ASLR} would fail as excepted if it is compiled with \texttt{-pie -fPIE}. 
As for the test cases reading a part of a function (\texttt{read-func})
or reading the content of a GOT (\texttt{read-GOT}),
no defense is detected on any of the platforms.

\subsubsection{Pointer integrity}
The test suite detects no defense against maliciously reading and modifying a code pointer.
Although compilers do warn about the arithmetic on a function pointer,
the test case still finishes without exception.
There is no defense found for protecting the VTable pointer as well.
The only type of defense is related to modifying an entry in the GOT.
Although GOT can be read on all platforms,
they are placed on read-only pages on Intel Xeon 8280, Jetson Nano and Raspberry Pi 4B
complying with the relocation read-only (RELRO) protection.
Note that most Linux distributions support partial RELRO by default.
For the platforms where \texttt{modify-GOT} succeeds,
the partial RELRO fails to cover the entry for the library function (\texttt{rand()}) hijacked.
The success of \texttt{modify-GOT} on HiFive Unleashed and Spike indicates
potential security issues that have not been properly resolved in the latest RISC-V compilers.

\subsubsection{Control-flow integrity}
It seems like most CFI related defenses are disabled by default on all platforms.
All tests that hijack the return address to a code exiting in memory succeed without exception.
Only the code injection tests are prevented by DEP as expected.
One extra benefit of using the latest RISC-V ISA (on Spike) is the failure of \texttt{\seqsplit{return-to-instruction-in-rodata}}
which tries to execute a code injected in constant data (read-only pages).
Starting from RISC-V privileged specification version 1.11~\cite{Waterman2019s},
read-only pages are no longer executable by default as a separate ``E'' flag is defined in the page table entry.
Since the privileged specification adopted by HiFive Unleashed is version 1.10,
such feature is not supported yet and the test succeeds just like on other platforms.
However, to our surprise, the test still passes on Spike when PIE is disabled (\texttt{-no-pie}).
Some parts of the global constant data might be put in the code section for a reduced memory footprint.

Similar to the tests for backward control-flow integrity,
all tests that hijack a function pointer succeed without a hitch except for the code injection tests.
The latest RISC-V provides the extra benefit of preventing code execution on read-only but non-execution pages.

The current situation on the protection of VTables are worrying.
It is found that 
not only can attackers replace a Vtable with another existing one as expected
but also fabricate a table in writable memory sections such as stack, heap and global data.
There is no check on whether a VTable is stored on read-only pages
although compilers are doing so to avoid direct modification.
The only defense that has been found is a protection against malicious reuse of VTable pointers enforced by the memory allocation algorithm.
The memory allocation algorithm does provide some protection against malicious VTable replacement.
According to the result of \texttt{call-wrong-func-vtable-released},
when an object is released, the VTable pointer is zeroed by the memory allocator.
Reusing a VTable pointer from a released object is therefore prevented.

\subsubsection{Summary of Results}
Comparing across platforms,
there is currently no significant difference for the provided security with default settings
while newer architectures (RISC-V) using the latest compiler and runtime libraries
show marginal security benefit thanks to the support of R$\oplus$X in page tables
and the protection against common heap-based UAF attacks implemented in the newer C library.

\subsection{Test with Different Flags and Compilers}

It is common that a defense is added to a compiler but disabled by default due to performance concerns.
As described in \emph{Q2} and \emph{Q3} of the introduction,
application designer may wish to explore 
their choices on different combination of compilers and defenses,
and security researchers would like to evaluate the strength of a certain defense on a specific platform.
The test suite would be a good tool for these kind of explorations.

For the investigation of different compiler flags,
we have compiled and installed the same version of the GCC compiler (version 10.3.0)
and the C library (GLIBC 2.32) on both Intel Xeon 8280 and Raspberry Pi 4B to cover Intel x86-64 and Arm AArch64.
No RISC-V platform is tested in this experiment because some important defenses,
such as the VTable verification (VTV)~\cite{Tice2014}, have not been ported to RISC-V yet.
We have also installed a latest LLVM 13 on Intel Xeon 8280 to explore the differences between compilers. 

\begin{table}[bt]
\footnotesize
\begin{center}
  \caption{Available sets of compiler flags}\label{tab:flags}
  \begin{tabular}{llc}
    \toprule
                 & compiler flags                            & compiler          \\
    \midrule
    Default      & -O2                                       & GCC and LLVM      \\
    \midrule
    RELRO        & -pie -fPIE -Wl,-z,relro,-z,now            & GCC and LLVM      \\
    Stack protection & -Wstack-protector -fstack-protector-all   & GCC and LLVM      \\
    VTV          & -fvtable-verify=std                       & GCC only          \\
    CFI          & -fvisibility=default -fsanitize=cfi       & LLVM only         \\
                 & -flto -fuse-ld=gold                       &                   \\
    \midrule
    All          & all the above                             & GCC and LLVM      \\
    ASan         & -fsanitize=address --param=asan-stack=1   & GCC and LLVM      \\
    None         & -z execstack -fno-stack-protector         & GCC and LLVM      \\
                 & (sysctl -w kernel.randomize\_va\_space=0) &                   \\
    \bottomrule
  \end{tabular}
\end{center}
\end{table}

After a small survey on the available security features in both compilers,
Table~\ref{tab:flags} lists the sets of compiler flags being explored:
\begin{itemize}
\item \textbf{Default}: Test using the latest compiler with the default compiler flags.
\item \textbf{RELRO} (RElocation Read-Only): Enforce the full protection of the GOT~\cite{HockeyInJune2011}.
\item \textbf{Stack protection}: Apply the stack smashing protection by inserting a canary on the stack~\cite{Cowan2003}.
\item \textbf{VTV}: Apply the GCC variant of defense against COOP attacks.
\item \textbf{CFI}: Similar to VTV, apply the LLVM variant of defense against all forward CFI attacks~\cite{Tice2014}.
\item \textbf{All}: Apply all the above defenses.
\item \textbf{ASan}: Enable the full address sanitizer at runtime.
\item \textbf{None}: Deliberately disable all proctections, including the one applied by the kernel, such as ASLR.
\end{itemize}

Some defenses are supported in compilers by not evaluated. 
Intel CET~\cite{CET2019} is supported by the latest compiler but the hardware is not ready yet.
Intel MPX~\cite{Ramakesavan2016} has been removed from GCC in recent years.
Arm PA~\cite{PALLVM2021} and Arm MTE~\cite{Bannister2019} are supported in LLVM
but unfortunately we do not have access to a compatible Arm platform for the test.
Stack clash protection has been tested but does not affect the result.

\begin{table}[bt]
\footnotesize
\begin{center}
  \caption{Test results with different flags and compilers}\label{tab:result-flags}
  \begin{tabular}{lccc}
    \toprule
                 & \rot{\shortstack[l]{Intel \\ Xeon 8280 \\ (GCC)}}
                         & \rot{\shortstack[l]{Raspberry \\ Pi 4B \\ (GCC)}}
                                 & \rot{\shortstack[l]{Intel \\ Xeon 8280 \\ (LLVM)}} \\
    \midrule
    Default      &  142  & 142   & 142 \\
    \midrule
    RELRO        &  141  & 141   & 140 \\
    Stack protection &  141  & 140   & 141 \\
    VTV          &  136  & 136   & N/A \\
    CFI          &  N/A  & N/A   & 141 \\
    \midrule
    All          &  134  & 133   & 138 \\
    ASan         &  8    & 8     & 21  \\
    None         &  155  & 147   & 154 \\
    \bottomrule
  \end{tabular}
\end{center}
\end{table}

Table~\ref{tab:result-flags} reveals the total number of passed test cases with different flags and compilers
while more detailed results are provide in Appendix~\ref{app:compiler-result}.

For the test of using different compiler flags,
it is found that enabling security features does defeat the targeted types of attacks
and there is no significant difference on the effectiveness of these features on different platforms,
at least for the platforms we have tested.
For the \textbf{default} setting using GCC,
Intel Xeon 8280 and Raspberry Pi 4B have exactly the same result of 142 passed tests.
Compared with the results in Table~\ref{tab:result-default},
the hijacking of GOT entry passes on both platforms as the full protection of GOT is deliberately disabled in the compilation of the compiler,
and four UAF on heap tests fail on Intel Xeon 8280 thanks to the newer C library.
By enabling the full GOT protection (case \textbf{RELRO}), the total number of tests passed reduces by one on both platforms.
Enabling \textbf{stack protection} has very limited protection on both platforms
because most ROP attacks can pinpoint the location of the return address and modify it without touching the canary.
Only the test trying to fake a call frame fails on both platform as it cannot create a correct canary.
As the only extra benefit, The test for return-to-libc~\cite{Pincus2004} fails on Raspberry Pi 4B
as the checking of canary in the function epilogue disturbs the malicious manipulation of function arguments.
\textbf{VTV} has exactly the same effect on both platforms that six COOP attacks have been stopped.
Any attack that hijacks a VTable to an injected one or even one failing the class hierarchy analysis~\cite{Jang2014}
is detected and stopped
but it is still possible to malicious reuse the VTables from relatives, such as the parent, child and sibling classes.
When \textbf{all} the mentioned security features are enabled,
the protection is simply combined together with 26 tests fail on Intel Xeon 8280
and 27 tests failed on Raspberry Pi 4B.
Address sanitizer (\textbf{ASan}) is a well known debugging tool that is normally disabled at production-phase.
We have also done an experiment to see its efficiency.
The result is impressive as only eight tests remain successful while all other 152 tests fail on both platforms.
The remaining tests include two access control tests (\texttt{read-func} and \texttt{read-GOT})
and six UAF attacks on stack.
We think the latter could be included by future address sanitizers.
Finally, when all protection features are deliberately disabled (the \textbf{none} case), including ASLR and DEP (execution on stack),
only five and thirteen tests fail on Intel Xeon 8280 and Raspberry Pi 4B respectively.
All the failed tests on Intel Xeon 8280 are heap-based UAF attacks thwarted by the latest C library.
Eight extra tests fails on Raspberry Pi 4B because code injection on heap and global data is still prohibited on Arm AArch64
even when execution on stack is allowed.
As a result, Arm AArch64 is marginally safer than Intel x86-64.
Nearly the same result is observed on the two RISC-V platforms excepted for \textbf{VTV} as it is not yet ported by GCC.

The test suite is portable enough that no change is needed when switching to LLVM using the \textbf{default} setting.
ASLR is found disabled as the distribution provided LLVM produces non-PIE executables by default
but arithmetic operations on function pointers are explicitly disallowed~\cite{Chen2019}.
The result is the same number of tests passed for the \textbf{default} case compared with GCC.
Enabling \textbf{RELRO} not only activates the full GOT protection but also ASLR as PIE is enabled.
The \textbf{stack protection} provided by LLVM has exactly the same effect as GCC.
Only the test trying to fake a call frame fails.
To our surprise, enabling \textbf{CFI} has almost no protection according to the result;
however, we believe the test suite in the current form cannot provide an accurate estimation for this protection.
The implementation of CFI protection on LLVM resorts to the link time optimization
and requires visibility to all definitions of virtual classes at link time.
This requirement forces the static linking of class definition mentioned in Section~\ref{subsec:cpp-opt} rather than dynamic linking through a shared library as in GCC.
As a result,
we suspect that all the illegal modifications on VTable pointers and function pointers are taken as valid operations by the link time analysis.
Consequently, enabling \textbf{all} the mentioned protection using LLVM results higher number of passed tests than GCC.
LLVM's address sanitizer (\textbf{ASan}) is seemly less effective than it on GCC.
It catches all ROP and COOP attacks but let go COP and JOP attacks.
It also fails to detect the modification on GOT entries.
On the good side, it catches nearly all UAF attacks,
including the UAF on stack attacks missed by GCC's ASan.
Finally, when all protections are disabled (case \textbf{none}),
LLVM and GCC have the same result except that the test applying arithmetic operations on function pointers fails to compile on LLVM.

\section{Discussion}\label{sec:discuss}

\subsection{Related Work}\label{sec:relate}


Computer architectures have for decades been constantly improved to satisfy our thirst for performance.
To quantitatively evaluate the performance gain brought by each improvement,
a number of test suites have been carefully designed and widely utilized.
LINPACK~\cite{Dongarra1979} was introduced in the 1970s to measure the computing power for numerical linear algebra and
is still used to rank today's supercomputers.
Dhrystone~\cite{Weicker1984} provides a suitable indicator for the general integer performance of computers.
CoreMark~\cite{GalOn2012} specializes in the performance of micro-controllers
while the more powerful general-purpose computers are normally compared using the SPEC benchmark suite~\cite{Henning2006}.
PARSEC~\cite{Bienia2008} concentrates on the performance of shared memory and multi-thread applications.



Using a test suite to measure the security of a certain system is not new either.
They began to appear in 2005,
when two papers~\cite{Kratkiewicz2005, Zhivich2005} proposed to use corpora of small and synthetic buffer overflow attacks (291 in \cite{Kratkiewicz2005} and 55 in \cite{Zhivich2005})
to test whether they can be detected by the then state-of-the-art software defenses, such as CRED~\cite{Ruwase2004} and CCured~\cite{Necula2005}.
Later in 2006, BASS~\cite{Poe2006} adopted an approach similar to the SPEC CPU test suite.
It contained seven relatively large test cases embedded with different types of vulnerabilities related to memory spatial safety.
It then provided a framework for the automatic generation of attacks exploiting the embedded vulnerabilities.
To our best knowledge, BASS was the first attempt for measuring the security of a computer platform
while its scope is limited to several types of memory spatial errors.
RIPE~\cite{Wilander2011} is probably the most utilized security test suite for defenses related to memory safety.
By exhausting the possible combinations of several attacking techniques, RIPE is able to cover 850 forms of buffer overflow and ROP attacks.
It has already been extensively utilized to verify the strength of various hardware supported defenses
against control-flow attacks~\cite{Kuznetsov2014,Song2016}.
However, the fact that it uses as many as 850 test cases to cover only buffer overflow and ROP attacks
indicates that reaching a full coverage by exhausting all possible combinations of vulnerabilities and attack techniques is unrealistic.
This partially motivates us to produce a new testing framework.

The design of security test suites has been revived in recent years.
The concept of a CPU security benchmark was introduced in \cite{Zhu2018}
which proposed several testing techniques also utilized in our test suite.
Unfortunately, the benchmark is not available for comparison,
no detail was revealed with regard to its internal structure,
and no result is provided in the paper.
CONFIRM~\cite{Xu2019} was a recently proposed test suite
to evaluate the compatibility and applicability of various control-flow integrity defenses
for different applications but it provides very little information on the security provided by the evaluated defenses.
Motivated by the lack of security evaluation by CONFIRM,
CBench~\cite{Li2020} was introduced to measure the gap between the practical security strength
and the claimed security of different control-flow integrity defenses.
It followed a similar approach with BASS.
A total of eighteen vulnerable programs were provided in seven categories.
By running the vulnerable programs and attacking them at runtime,
CBench measures the security level of the applied CFI defense.
Compared with the test suite proposed in this paper,
CBench experiments with full-fledged attacks while overlooking the relationship between individual vulnerabilities. 
It concentrates on evaluating the strength of the defense mechanisms themselves but not the platform implementing them.
Consequently, CBench does not need to support multiple platforms and runs only on x86-64.

\subsection{Future Work}

The testing framework proposed in this paper is far from finished.
It would need some significant extensions to become
a full-fledged test suite measuring the security supported by the processor hardware.

\subsubsection{Missing test cases for memory safety}
One category of vulnerabilities missing in the current test suite is related to data racing in multi-threaded applications~\cite{Li2020}.
When the integrity check on a pointer and the following pointer dereferencing do not operate atomically,
an attacker running on a parallel thread
might hijack the pointer just after the integrity check but before the pointer dereferencing.
Currently, all test cases are single threaded.
Supporting multi-threaded tests for data racing conditions is one of our immediate future works.

Test cases for DOP attacks~\cite{Hu2016} are being gradually added.
The number of test cases for type confusion and malicious initialization of uninitiated variables
will be substantially increased to cover all possibilities.
It is also reported that a significant portion of DOP attacks violate the visibility scope of variables~\cite{Nyman2019}.
Applying fine-grained compartmentalization of data memory to enforce the visibility scope of variables
is a promising defense against DOP attacks and is not yet fully covered by the test suite.

\subsubsection{Missing test cases for other types of safety}
Although important, memory safety is only one of the many categories of security that can be supported by the processor hardware.
At least two other categories of defenses should be included in this test suite:
hardware supported defenses against the \emph{software-based side-channel attacks} and the \emph{transient attacks}~\cite{Canella2019}.

The software-based side-channel attacks include all side-channel or covert channel attacks that can be launched purely by software.
They include both the conflict-based~\cite{Yan2017} and flush-based~\cite{Yarom2014} cache side-channel attacks.
They contain also the attacks causing timing channels inside the processor core by maliciously competing resources,
such as the recent port smash attack~\cite{Aldaya2019}.
Although cache side-channel attacks against the traditional level-one caches are thoroughly evaluated by \cite{Deng2020},
this test suite has not covered the more prevalent attacks against the last-level caches,
cannot be easily ported to other non-x86 architectures,
and still misses coverage on the new variants of side-channel attacks~\cite{Bourgeat2020, Purnal2021, Song2021}
targeting the recently revived randomized caches~\cite{Qureshi2019, Werner2019}.

The term transient execution attack was first introduced in \cite{Canella2019} and used to summarize all attacks that
exploit the speculative execution in modern out-of-order processors.
It includes all variants of the Meltdown~\cite{Lipp2018} and the Spectra~\cite{Kocher2019} attacks.
New variants of transient execution attacks~\cite{Bulck2018, Canella2019a, Bulck2020}
are still being found in recent years along with various new defense proposals~\cite{Saileshwar2019, Ainsworth2020, Loughlin2021}.
A proper test suite to systematically evaluate the remaining attack vectors left by the different defenses is urgently needed.

\subsubsection{Vulnerability scoring system}
We still need a vulnerability scoring system to interpret the test result and
translate it into a score quantitively indicating the security supported by the processor.

Every vulnerability recorded in the common vulnerabilities and exposures (CVE) database~\cite{CVE1999} is assigned with a vulnerability score
measured by the common vulnerability scoring system (CVSS)~\cite{CVSS}.
However, we cannot directly adopt this scoring system for evaluating architectural vulnerabilities
as the metric used is strictly software oriented.
We have also investigated the common weakness enumeration (CWE)~\cite{CWE} and found 58 related CWEs.
By averaging the CVSS scores of the CVEs related to each CWE,
we might be able to calculate an equivalent score for each test case using the existing scores from CVEs.
Unfortunately, most of the 58 CWEs are linked to the test cases regarding to spatial and temporal safety of memory,
while most of the control flow related test cases have no connected CWE.
As a result, we cannot directly borrow the existing CVSS to score the architectural vulnerabilities.
This is another open challenge that requires solutions.

\section{Conclusion}\label{sec:con}

A comprehensive and cross-platform test suite has been produced to measure memory safety.
It is also our first step in producing a fully-fledged test suite to measure the security supported by the processor hardware.
The initial test suite currently contains 160 test cases covering
spatial and temporal safety of memory, memory access control,
pointer integrity and control-flow integrity.
Each type of vulnerabilities and their related defenses
have been individually evaluated by one or more test cases.
To show its usefulness,
the test suite has been ported to six platforms using three different ISAs,
including Intel x86-64, Arm AArch64 and RISC-V.
According to our experiment results,
most memory safety vulnerabilities are still exploitable with the default compiler settings.
Enabling extra security features in compilers does defeat the targeted types of attacks.
Although address sanitizer is often a tool for debugging and not a practical production-phase protection measure, it is indeed effective in catching attacks.
Comparing across platforms, newer architectures using the latest compiler and runtime libraries show marginal security benefit, at least for the platforms we have tested.
Comparing between compilers, LLVM has similar security support with GCC
but its address sanitizer is seemly weaker.



\section*{Availability}

The benchmark has been open sourced on GitHub at \\
\url{https://github.com/comparch-security/cpu-sec-bench}.

\section*{Acknowledgment}

Yuhui Zhang has contributed the early versions of this benchmark.
The HiFive Unleashed board was kindly borrowed from Xiongfei Guo.
This work was partially supported by the National Natural Science Foundation of China under grant No. 61802402,
the CAS Pioneer Hundred Talents Program,
and internal grants from the Institute of Information Engineering, CAS.
Any opinions, findings, conclusions, and recommendations expressed in this paper
are those of the authors and do not necessarily reflect the views of the funding parties.

\bibliographystyle{ACM-Reference-Format}
\bibliography{reference}

\newpage
\appendix

\newcommand*\REX{$\star$}   
\newcommand*\RFA{$\times$}  
\newcommand*\RNT{$\otimes$} 
\newcommand*\RPA{$\surd$}   
\newcommand*\RNA{$-$}       

\section{Results for Different Architectures}\label{app:mem-safety}

\RPA: test passes (no defense detected);\\
\RFA: test failed;\\
\REX: exception raised;\\
\RNT: not tested due to failed prerequisites.

\subsection{Spatial Safety}
\begin{center}
  \footnotesize
\begin{tabular}{lcccccc}
    \toprule
    & \rot{\MMlizz} & \rot{\MMstockport} & \rot{\MMJetson} & \rot{\MMRasp} &  \rot{\MMHiFive} & \rot{\MMSpike} \\
    \midrule
    overflow-read-index-stack                & \RPA & \RPA & \RPA & \RPA & \RPA & \RPA \\
    overflow-read-index-heap                 & \RPA & \RPA & \RPA & \RPA & \RPA & \RPA \\
    overflow-read-index-data                 & \RPA & \RPA & \RPA & \RPA & \RPA & \RPA \\
    overflow-read-index-rodata               & \RPA & \RPA & \RPA & \RPA & \RPA & \RPA \\
    overflow-read-ptr-stack                  & \RPA & \RPA & \RPA & \RPA & \RPA & \RPA \\
    overflow-read-ptr-heap                   & \RPA & \RPA & \RPA & \RPA & \RPA & \RPA \\
    overflow-read-ptr-data                   & \RPA & \RPA & \RPA & \RPA & \RPA & \RPA \\
\bottomrule
  \end{tabular}
\end{center}

\begin{center}
  \footnotesize
  \begin{tabular}{lcccccc}
    \toprule
    & \rot{\MMlizz} & \rot{\MMstockport} & \rot{\MMJetson} & \rot{\MMRasp} &  \rot{\MMHiFive} & \rot{\MMSpike} \\
    \midrule
    read-cross-section-stack-to-heap-index   & \RPA & \RPA & \RPA & \RPA & \RPA & \RPA \\
    read-cross-section-stack-to-data-index   & \RPA & \RPA & \RPA & \RPA & \RPA & \RPA \\
    read-cross-section-stack-to-rodata-index & \RPA & \RPA & \RPA & \RPA & \RPA & \RPA \\
    read-cross-section-heap-to-stack-index   & \RPA & \RPA & \RPA & \RPA & \RPA & \RPA \\
    read-cross-section-heap-to-data-index    & \RPA & \RPA & \RPA & \RPA & \RPA & \RPA \\
    read-cross-section-heap-to-rodata-index  & \RPA & \RPA & \RPA & \RPA & \RPA & \RPA \\
    read-cross-section-data-to-stack-index   & \RPA & \RPA & \RPA & \RPA & \RPA & \RPA \\
\bottomrule
  \end{tabular}
\end{center}

\subsection{Temporal Safety}
\begin{center}
  \footnotesize
  \begin{tabular}{lcccccc}
    \toprule
    & \rot{\MMlizz} & \rot{\MMstockport} & \rot{\MMJetson} & \rot{\MMRasp} &  \rot{\MMHiFive} & \rot{\MMSpike} \\
    \midrule
    read-after-free-org-heap    & \RPA & \RPA & \RPA & \RFA & \RFA & \RFA \\
    read-after-free-alias-heap  & \RPA & \RPA & \RPA & \RFA & \RFA & \RFA \\
    write-after-free-heap       & \RPA & \RPA & \RPA & \RPA & \RPA & \RPA \\
    \midrule
    reallocate-heap             & \RPA & \RPA & \RPA & \RPA & \RPA & \RPA \\
    \midrule
    access-after-reclaim-heap   & \RPA & \RPA & \RPA & \RFA & \RFA & \RFA \\
    write-before-reclaim-heap   & \RPA & \RPA & \RPA & \RFA & \RFA & \RFA \\
    write-after-reclaim-heap    & \RPA & \RPA & \RPA & \RPA & \RPA & \RPA \\
    \midrule
    read-after-free-alias-stack & \RPA & \RPA & \RPA & \RPA & \RPA & \RPA \\
    write-after-free-stack      & \RPA & \RPA & \RPA & \RPA & \RPA & \RPA \\
    \midrule
    reallocate-stack            & \RPA & \RPA &\RPA  & \RPA & \RPA & \RPA \\
    \midrule
    access-after-reclaim-stack  & \RPA & \RPA & \RPA & \RPA & \RPA & \RPA \\
    write-before-reclaim-stack  & \RPA & \RPA & \RPA & \RPA & \RPA & \RPA \\
    write-after-reclaim-stack   & \RPA & \RPA & \RPA & \RPA & \RPA & \RPA \\
    \bottomrule
  \end{tabular}
\end{center}

\subsection{Access Control}
\begin{center}
  \footnotesize
  \begin{tabular}{lcccccc}
    \toprule
    & \rot{\MMlizz} & \rot{\MMstockport} & \rot{\MMJetson} & \rot{\MMRasp} &  \rot{\MMHiFive} & \rot{\MMSpike} \\
    \midrule
    check-ASLR                  & \RPA & \RFA & \RFA & \RFA & \RFA & \RPA \\
    read-func                   & \RPA & \RPA & \RPA & \RPA & \RPA & \RPA \\
    read-GOT                    & \RPA & \RPA & \RPA & \RPA & \RPA & \RPA \\
    \bottomrule
  \end{tabular}
\end{center}

\subsection{Pointer Integrity}

\begin{center}
  \footnotesize
  \begin{tabular}{lcccccc}
    \toprule
    & \rot{\MMlizz} & \rot{\MMstockport} & \rot{\MMJetson} & \rot{\MMRasp} &  \rot{\MMHiFive} & \rot{\MMSpike} \\
    \midrule
    func-pointer-assign     & \RPA & \RPA & \RPA & \RPA & \RPA & \RPA \\
    func-pointer-arithmetic & \RPA & \RPA & \RPA & \RPA & \RPA & \RPA \\
    \midrule
    read-vtable-pointer     & \RPA & \RPA & \RPA & \RPA & \RPA & \RPA \\
    write-vtable-pointer    & \RPA & \RPA & \RPA & \RPA & \RPA & \RPA \\
    \midrule
    modify-GOT              & \RPA & \REX & \REX & \REX & \RPA & \RPA \\
    \bottomrule
  \end{tabular}
\end{center}

\subsection{Control-Flow Integrity}

\begin{center}
  \footnotesize
  \begin{tabular}{lcccccc}
    \toprule
    & \rot{\MMlizz} & \rot{\MMstockport} & \rot{\MMJetson} & \rot{\MMRasp} &  \rot{\MMHiFive} & \rot{\MMSpike} \\
    \midrule
    return-to-instruction-in-stack                   & \REX & \REX & \REX & \REX & \REX & \REX \\
    return-to-instruction-in-heap                    & \REX & \REX & \REX & \REX & \REX & \REX \\
    return-to-instruction-in-data                    & \REX & \REX & \REX & \REX & \REX & \REX \\
    return-to-instruction-in-rodata                  & \RPA & \RPA & \RPA & \RPA & \RPA & \REX \\
    \midrule
    return-to-wrong-call-site-within-static-analysis & \RPA & \RPA & \RPA & \RPA & \RPA & \RPA \\
    return-to-wrong-call-site                        & \RPA & \RPA & \RPA & \RPA & \RPA & \RPA \\
    return-to-non-call-site                          & \RPA & \RPA & \RPA & \RPA & \RPA & \RPA \\
\bottomrule
  \end{tabular}
\end{center}

\begin{center}
  \footnotesize
  \begin{tabular}{lcccccc}
    \toprule
    & \rot{\MMlizz} & \rot{\MMstockport} & \rot{\MMJetson} & \rot{\MMRasp} &  \rot{\MMHiFive} & \rot{\MMSpike} \\
    \midrule
    call-wrong-func-vtable-offset          & \RPA & \RPA & \RPA & \RPA & \RPA & \RPA \\
    call-wrong-num-arg-vtable              & \RPA & \RPA & \RPA & \RPA & \RPA & \RPA \\
    \midrule
    jump-instruction-in-stack              & \REX & \REX & \REX & \REX & \REX & \REX \\
    jump-instruction-in-heap               & \REX & \REX & \REX & \REX & \REX & \REX \\
    jump-instruction-in-data               & \REX & \REX & \REX & \REX & \REX & \REX \\
    jump-instruction-in-rodata             & \RPA & \RPA & \RPA & \RPA & \RPA & \REX \\
    jump-mid-func                          & \RPA & \RPA & \RPA & \RPA & \RPA & \RPA \\
    \bottomrule
  \end{tabular}
\end{center}

\section{Results for Different Compiler Flags}\label{app:compiler-result}

\subsection{GCC on Intel Xeon 8280}

\subsubsection{Summary}
\begin{center}
  \footnotesize
  \begin{tabular}{lccccccc}
    \toprule
    & \rot{Default} & \rot{RELRO} & \rot{\shortstack[l]{Stack protection}} & \rot{VTV} &  \rot{All} & \rot{ASan} & \rot{None}\\
    \midrule
    Spatial safety (98)     & 98   & 98     & 98      & 98      & 98      & \bf{0}  & 98 \\
    Temporal safety (13)    & 9    & 9      & 9       & 9       & 9       & \bf{6}  & 9  \\
    Access control (3)      & 2    & 2      & 2       & 2       & 2       & 2       & \bf{3}\\
    Pointer integrity (5)   & 5    & \bf{4} & 5       & 5       & \bf{4}  & \bf{0}  & 5 \\
    CFI (41)                & 28   & 28     & \bf{27} & \bf{22} & \bf{21} & \bf{0}  & \bf{40} \\
    \midrule
    Total (160)             & 142  & 141    & 141     & 136     & 134     & 8       & 155 \\
    \bottomrule
  \end{tabular}
\end{center}

\subsubsection{Spatial Safety}
\begin{center}
  \footnotesize
  \begin{tabular}{lccccccc}
    \toprule
    & \rot{Default} & \rot{RELRO} & \rot{\shortstack[l]{Stack protection}} & \rot{VTV} &  \rot{All} & \rot{ASan} & \rot{None}\\
    \midrule
    overflow-read-index-stack                & \RPA & \RPA & \RPA & \RPA & \RPA & \RFA & \RPA \\
    overflow-read-index-heap                 & \RPA & \RPA & \RPA & \RPA & \RPA & \RFA & \RPA \\
    overflow-read-index-data                 & \RPA & \RPA & \RPA & \RPA & \RPA & \RFA & \RPA \\
    overflow-read-index-rodata               & \RPA & \RPA & \RPA & \RPA & \RPA & \RFA & \RPA \\
    overflow-read-ptr-stack                  & \RPA & \RPA & \RPA & \RPA & \RPA & \RFA & \RPA \\
    overflow-read-ptr-heap                   & \RPA & \RPA & \RPA & \RPA & \RPA & \RFA & \RPA \\
    overflow-read-ptr-data                   & \RPA & \RPA & \RPA & \RPA & \RPA & \RFA & \RPA \\
\bottomrule
  \end{tabular}
\end{center}

\begin{center}
  \footnotesize
  \begin{tabular}{lccccccc}
    \toprule
    & \rot{Default} & \rot{RELRO} & \rot{\shortstack[l]{Stack protection}} & \rot{VTV} &  \rot{All} & \rot{ASan} & \rot{None}\\
    \midrule
    read-cross-object-index-stack            & \RPA & \RPA & \RPA & \RPA & \RPA & \RNT & \RPA \\
    read-cross-object-index-heap             & \RPA & \RPA & \RPA & \RPA & \RPA & \RNT & \RPA \\
    read-cross-object-index-data             & \RPA & \RPA & \RPA & \RPA & \RPA & \RNT & \RPA \\
    read-cross-object-index-rodata           & \RPA & \RPA & \RPA & \RPA & \RPA & \RNT & \RPA \\
    read-cross-object-ptr-stack              & \RPA & \RPA & \RPA & \RPA & \RPA & \RNT & \RPA \\
    read-cross-object-ptr-heap               & \RPA & \RPA & \RPA & \RPA & \RPA & \RNT & \RPA \\
    read-cross-object-ptr-data               & \RPA & \RPA & \RPA & \RPA & \RPA & \RNT & \RPA \\
\bottomrule
  \end{tabular}
\end{center}

\begin{center}
  \footnotesize
  \begin{tabular}{lccccccc}
    \toprule
    & \rot{Default} & \rot{RELRO} & \rot{\shortstack[l]{Stack protection}} & \rot{VTV} &  \rot{All} & \rot{ASan} & \rot{None}\\
    \midrule
    spray-cross-page-stack                   & \RPA & \RPA & \RPA & \RPA & \RPA & \RNT & \RPA \\
    spray-cross-page-heap                    & \RPA & \RPA & \RPA & \RPA & \RPA & \RNT & \RPA \\
    \bottomrule
  \end{tabular}
\end{center}

\subsubsection{Temporal Safety}
\begin{center}
  \footnotesize
  \begin{tabular}{lccccccc}
    \toprule
    & \rot{Default} & \rot{RELRO} & \rot{\shortstack[l]{Stack protection}} & \rot{VTV} &  \rot{All} & \rot{ASan} & \rot{None}\\
    \midrule
    read-after-free-org-heap    & \RFA & \RFA & \RFA & \RFA & \RFA & \RFA & \RFA \\
    read-after-free-alias-heap  & \RFA & \RFA & \RFA & \RFA & \RFA & \RFA & \RFA \\
    write-after-free-heap       & \RPA & \RPA & \RPA & \RPA & \RPA & \RFA & \RPA \\
    \midrule
    reallocate-heap             & \RPA & \RPA & \RPA & \RPA & \RPA & \RFA & \RPA \\
    \midrule
    access-after-reclaim-heap   & \RFA & \RFA & \RFA & \RFA & \RFA & \RNT & \RFA \\
    write-before-reclaim-heap   & \RFA & \RFA & \RFA & \RFA & \RFA & \RNT & \RFA \\
    write-after-reclaim-heap    & \RPA & \RPA & \RPA & \RPA & \RPA & \RNT & \RPA \\
    \midrule
    read-after-free-alias-stack & \RPA & \RPA & \RPA & \RPA & \RPA & \RPA & \RPA \\
    write-after-free-stack      & \RPA & \RPA & \RPA & \RPA & \RPA & \RPA & \RPA \\
    \midrule
    reallocate-stack            & \RPA & \RPA &\RPA  & \RPA & \RPA & \RPA & \RPA \\
    \midrule
    access-after-reclaim-stack  & \RPA & \RPA & \RPA & \RPA & \RPA & \RPA & \RPA \\
    write-before-reclaim-stack  & \RPA & \RPA & \RPA & \RPA & \RPA & \RPA & \RPA \\
    write-after-reclaim-stack   & \RPA & \RPA & \RPA & \RPA & \RPA & \RPA & \RPA \\
    \bottomrule
  \end{tabular}
\end{center}

\subsubsection{Access Control}
\begin{center}
  \footnotesize
  \begin{tabular}{lccccccc}
    \toprule
    & \rot{Default} & \rot{RELRO} & \rot{\shortstack[l]{Stack protection}} & \rot{VTV} &  \rot{All} & \rot{ASan} & \rot{None}\\
    \midrule
    check-ASLR                  & \RFA & \RFA & \RFA & \RFA & \RFA & \RFA& \RPA \\
    read-func                   & \RPA & \RPA & \RPA & \RPA & \RPA & \RPA& \RPA \\
    read-GOT                    & \RPA & \RPA & \RPA & \RPA & \RPA & \RPA& \RPA \\
    \bottomrule
  \end{tabular}
\end{center}

\subsubsection{Pointer Integrity}

\begin{center}
  \footnotesize
  \begin{tabular}{lccccccc}
    \toprule
    & \rot{Default} & \rot{RELRO} & \rot{\shortstack[l]{Stack protection}} & \rot{VTV} &  \rot{All} & \rot{ASan} & \rot{None}\\
    \midrule
    func-pointer-assign     & \RPA & \RPA & \RPA & \RPA & \RPA & \RFA & \RPA \\
    func-pointer-arithmetic & \RPA & \RPA & \RPA & \RPA & \RPA & \RNT & \RPA \\
    \midrule
    read-vtable-pointer     & \RPA & \RPA & \RPA & \RPA & \RPA & \RFA & \RPA \\
    write-vtable-pointer    & \RPA & \RPA & \RPA & \RPA & \RPA & \RNT & \RPA \\
    \midrule
    modify-GOT              & \RPA & \REX & \RPA & \RPA & \REX & \RFA & \RPA \\
    \bottomrule
  \end{tabular}
\end{center}

\subsubsection{Control-Flow Integrity}

\begin{center}
  \footnotesize
  \begin{tabular}{lccccccc}
    \toprule
    & \rot{Default} & \rot{RELRO} & \rot{\shortstack[l]{Stack protection}} & \rot{VTV} &  \rot{All} & \rot{ASan} & \rot{None}\\
    \midrule
    return-to-instruction-in-stack                   & \REX & \REX & \REX & \REX & \REX & \RNT & \RPA \\
    return-to-instruction-in-heap                    & \REX & \REX & \REX & \REX & \REX & \RNT & \RPA \\
    return-to-instruction-in-data                    & \REX & \REX & \REX & \REX & \REX & \RNT & \RPA \\
    return-to-instruction-in-rodata                  & \RPA & \RPA & \RPA & \RPA & \RPA & \RNT & \RPA \\
    \bottomrule
  \end{tabular}
\end{center}

\begin{center}
  \footnotesize
  \begin{tabular}{lccccccc}
    \toprule
    & \rot{Default} & \rot{RELRO} & \rot{\shortstack[l]{Stack protection}} & \rot{VTV} &  \rot{All} & \rot{ASan} & \rot{None}\\
    \midrule
    return-to-wrong-call-site-within-static-analysis & \RPA & \RPA & \RPA & \RPA & \RPA & \RNT & \RPA \\
    return-to-wrong-call-site                        & \RPA & \RPA & \RPA & \RPA & \RPA & \RNT & \RPA \\
    return-to-non-call-site                          & \RPA & \RPA & \RPA & \RPA & \RPA & \RNT & \RPA \\
    return-to-func                                   & \RPA & \RPA & \RPA & \RPA & \RPA & \RNT & \RPA \\
    return-to-libc                                   & \RPA & \RPA & \RPA & \RPA & \RPA & \RNT & \RPA \\
    return-without-call                              & \RPA & \RPA & \REX & \RPA & \REX & \RNT & \RPA \\
    \midrule
    call-instruction-in-stack              & \REX & \REX & \REX & \REX & \REX & \RFA & \RPA \\
\bottomrule
  \end{tabular}
\end{center}

\subsection{GCC on Raspberry Pi 4B}

\subsubsection{Summary}
\begin{center}
  \footnotesize
  \begin{tabular}{lccccccc}
    \toprule
    & \rot{Default} & \rot{RELRO} & \rot{\shortstack[l]{Stack protection}} & \rot{VTV} &  \rot{All} & \rot{ASan} & \rot{None}\\
    \midrule
    Spatial safety (98)     & 98   & 98     & 98      & 98      & 98      & \bf{0}  & 98 \\
    Temporal safety (13)    & 9    & 9      & 9       & 9       & 9       & \bf{6}  & 9  \\
    Access control (3)      & 2    & 2      & 2       & 2       & 2       & 2       & \bf{3}\\
    Pointer integrity (5)   & 5    & \bf{4} & 5       & 5       & \bf{4}  & \bf{0}  & 5 \\
    CFI (41)                & 28   & 28     & \bf{26} & \bf{22} & \bf{20} & \bf{0}  & \bf{32} \\
    \midrule
    Total (160)             & 142  & 141    & 140     & 136     & 133     & 8       & 147 \\
    \bottomrule
  \end{tabular}
\end{center}

\newpage
\subsubsection{Spatial Safety}
\begin{center}
  \footnotesize
  \begin{tabular}{lccccccc}
    \toprule
    & \rot{Default} & \rot{RELRO} & \rot{\shortstack[l]{Stack protection}} & \rot{VTV} &  \rot{All} & \rot{ASan} & \rot{None}\\
    \midrule
    overflow-read-index-stack                & \RPA & \RPA & \RPA & \RPA & \RPA & \RFA & \RPA \\
    overflow-read-index-heap                 & \RPA & \RPA & \RPA & \RPA & \RPA & \RFA & \RPA \\
    overflow-read-index-data                 & \RPA & \RPA & \RPA & \RPA & \RPA & \RFA & \RPA \\
    overflow-read-index-rodata               & \RPA & \RPA & \RPA & \RPA & \RPA & \RFA & \RPA \\
    overflow-read-ptr-stack                  & \RPA & \RPA & \RPA & \RPA & \RPA & \RFA & \RPA \\
    overflow-read-ptr-heap                   & \RPA & \RPA & \RPA & \RPA & \RPA & \RFA & \RPA \\
    overflow-read-ptr-data                   & \RPA & \RPA & \RPA & \RPA & \RPA & \RFA & \RPA \\
\bottomrule
  \end{tabular}
\end{center}

\begin{center}
  \footnotesize
  \begin{tabular}{lccccccc}
    \toprule
    & \rot{Default} & \rot{RELRO} & \rot{\shortstack[l]{Stack protection}} & \rot{VTV} &  \rot{All} & \rot{ASan} & \rot{None}\\
    \midrule
    read-cross-section-rodata-to-stack-index & \RPA & \RPA & \RPA & \RPA & \RPA & \RNT & \RPA \\
    read-cross-section-rodata-to-heap-index  & \RPA & \RPA & \RPA & \RPA & \RPA & \RNT & \RPA \\
    read-cross-section-rodata-to-data-index  & \RPA & \RPA & \RPA & \RPA & \RPA & \RNT & \RPA \\
    read-cross-section-stack-to-heap-ptr     & \RPA & \RPA & \RPA & \RPA & \RPA & \RNT & \RPA \\
    read-cross-section-stack-to-data-ptr     & \RPA & \RPA & \RPA & \RPA & \RPA & \RNT & \RPA \\
    read-cross-section-stack-to-rodata-ptr   & \RPA & \RPA & \RPA & \RPA & \RPA & \RNT & \RPA \\
    read-cross-section-heap-to-stack-ptr     & \RPA & \RPA & \RPA & \RPA & \RPA & \RNT & \RPA \\
\bottomrule
  \end{tabular}
\end{center}

\subsubsection{Temporal Safety}
\begin{center}
  \footnotesize
  \begin{tabular}{lccccccc}
    \toprule
    & \rot{Default} & \rot{RELRO} & \rot{\shortstack[l]{Stack protection}} & \rot{VTV} &  \rot{All} & \rot{ASan} & \rot{None}\\
    \midrule
    read-after-free-org-heap    & \RFA & \RFA & \RFA & \RFA & \RFA & \RFA & \RFA \\
    read-after-free-alias-heap  & \RFA & \RFA & \RFA & \RFA & \RFA & \RFA & \RFA \\
    write-after-free-heap       & \RPA & \RPA & \RPA & \RPA & \RPA & \RFA & \RPA \\
    \midrule
    reallocate-heap             & \RPA & \RPA & \RPA & \RPA & \RPA & \RFA & \RPA \\
    \midrule
    access-after-reclaim-heap   & \RFA & \RFA & \RFA & \RFA & \RFA & \RNT & \RFA \\
    write-before-reclaim-heap   & \RFA & \RFA & \RFA & \RFA & \RFA & \RNT & \RFA \\
    write-after-reclaim-heap    & \RPA & \RPA & \RPA & \RPA & \RPA & \RNT & \RPA \\
    \midrule
    read-after-free-alias-stack & \RPA & \RPA & \RPA & \RPA & \RPA & \RPA & \RPA \\
    write-after-free-stack      & \RPA & \RPA & \RPA & \RPA & \RPA & \RPA & \RPA \\
    \midrule
    reallocate-stack            & \RPA & \RPA &\RPA  & \RPA & \RPA & \RPA & \RPA \\
    \midrule
    access-after-reclaim-stack  & \RPA & \RPA & \RPA & \RPA & \RPA & \RPA & \RPA \\
    write-before-reclaim-stack  & \RPA & \RPA & \RPA & \RPA & \RPA & \RPA & \RPA \\
    write-after-reclaim-stack   & \RPA & \RPA & \RPA & \RPA & \RPA & \RPA & \RPA \\
    \bottomrule
  \end{tabular}
\end{center}

\newpage
\subsubsection{Access Control}
\begin{center}
  \footnotesize
  \begin{tabular}{lccccccc}
    \toprule
    & \rot{Default} & \rot{RELRO} & \rot{\shortstack[l]{Stack protection}} & \rot{VTV} &  \rot{All} & \rot{ASan} & \rot{None}\\
    \midrule
    check-ASLR                  & \RFA & \RFA & \RFA & \RFA & \RFA & \RFA& \RPA \\
    read-func                   & \RPA & \RPA & \RPA & \RPA & \RPA & \RPA& \RPA \\
    read-GOT                    & \RPA & \RPA & \RPA & \RPA & \RPA & \RPA& \RPA \\
    \bottomrule
  \end{tabular}
\end{center}

\subsubsection{Pointer Integrity}

\begin{center}
  \footnotesize
  \begin{tabular}{lccccccc}
    \toprule
    & \rot{Default} & \rot{RELRO} & \rot{\shortstack[l]{Stack protection}} & \rot{VTV} &  \rot{All} & \rot{ASan} & \rot{None}\\
    \midrule
    func-pointer-assign     & \RPA & \RPA & \RPA & \RPA & \RPA & \RFA & \RPA \\
    func-pointer-arithmetic & \RPA & \RPA & \RPA & \RPA & \RPA & \RNT & \RPA \\
    \midrule
    read-vtable-pointer     & \RPA & \RPA & \RPA & \RPA & \RPA & \RFA & \RPA \\
    write-vtable-pointer    & \RPA & \RPA & \RPA & \RPA & \RPA & \RNT & \RPA \\
    \midrule
    modify-GOT              & \RPA & \REX & \RPA & \RPA & \REX & \RFA & \RPA \\
    \bottomrule
  \end{tabular}
\end{center}

\subsubsection{Control-Flow Integrity}

\begin{center}
  \footnotesize
  \begin{tabular}{lccccccc}
    \toprule
    & \rot{Default} & \rot{RELRO} & \rot{\shortstack[l]{Stack protection}} & \rot{VTV} &  \rot{All} & \rot{ASan} & \rot{None}\\
    \midrule
    return-to-instruction-in-stack                   & \REX & \REX & \REX & \REX & \REX & \RNT & \RPA \\
    return-to-instruction-in-heap                    & \REX & \REX & \REX & \REX & \REX & \RNT & \REX \\
    return-to-instruction-in-data                    & \REX & \REX & \REX & \REX & \REX & \RNT & \REX \\
    return-to-instruction-in-rodata                  & \RPA & \RPA & \RPA & \RPA & \RPA & \RNT & \RPA \\
    \midrule
    return-to-wrong-call-site-within-static-analysis & \RPA & \RPA & \RPA & \RPA & \RPA & \RNT & \RPA \\
    return-to-wrong-call-site                        & \RPA & \RPA & \RPA & \RPA & \RPA & \RNT & \RPA \\
    return-to-non-call-site                          & \RPA & \RPA & \RPA & \RPA & \RPA & \RNT & \RPA \\
\bottomrule
  \end{tabular}
\end{center}

\begin{center}
  \footnotesize
  \begin{tabular}{lccccccc}
    \toprule
    & \rot{Default} & \rot{RELRO} & \rot{\shortstack[l]{Stack protection}} & \rot{VTV} &  \rot{All} & \rot{ASan} & \rot{None}\\
    \midrule
    jump-instruction-in-stack              & \REX & \REX & \REX & \REX & \REX & \RNT & \RPA \\
    jump-instruction-in-heap               & \REX & \REX & \REX & \REX & \REX & \RNT & \REX \\
    jump-instruction-in-data               & \REX & \REX & \REX & \REX & \REX & \RNT & \REX \\
    jump-instruction-in-rodata             & \RPA & \RPA & \RPA & \RPA & \RPA & \RNT & \RPA \\
    jump-mid-func                          & \RPA & \RPA & \RPA & \RPA & \RPA & \RFA & \RPA \\
    \bottomrule
  \end{tabular}
\end{center}

\subsection{LLVM on Intel Xeon 8280}

\subsubsection{Summary}
\begin{center}
  \footnotesize
  \begin{tabular}{lccccccc}
    \toprule
    & \rot{Default} & \rot{RELRO} & \rot{\shortstack[l]{Stack protection}} & \rot{CFI} &  \rot{All} & \rot{ASan} & \rot{None}\\
    \midrule
    Spatial safety (98)     & 98   & 98     & 98      & 98      & 98      & \bf{0}  & 98 \\
    Temporal safety (13)    & 9    & 9      & 9       & 9       & 9       & \bf{4}  & 9  \\
    Access control (3)      & 3    & \bf{2} & 3       & 3       & \bf{2}  & 3       & 3  \\
    Pointer integrity (5)   & 4    & \bf{3} & 4       & 4       & \bf{3}  & \bf{2}  & 4  \\
    CFI (41)                & 28   & 28     & \bf{27} & \bf{27} & \bf{26} & \bf{12} & \bf{40} \\
    \midrule
    Total (160)             & 142  & 140    & 141     & 141     & 138     & 21      & 154 \\
    \bottomrule
  \end{tabular}
\end{center}

\subsubsection{Spatial Safety}
\begin{center}
  \footnotesize
  \begin{tabular}{lccccccc}
    \toprule
    & \rot{Default} & \rot{RELRO} & \rot{\shortstack[l]{Stack protection}} & \rot{CFI} &  \rot{All} & \rot{ASan} & \rot{None}\\
    \midrule
    overflow-read-index-stack                & \RPA & \RPA & \RPA & \RPA & \RPA & \RFA & \RPA \\
    overflow-read-index-heap                 & \RPA & \RPA & \RPA & \RPA & \RPA & \RFA & \RPA \\
    overflow-read-index-data                 & \RPA & \RPA & \RPA & \RPA & \RPA & \RFA & \RPA \\
    overflow-read-index-rodata               & \RPA & \RPA & \RPA & \RPA & \RPA & \RFA & \RPA \\
    overflow-read-ptr-stack                  & \RPA & \RPA & \RPA & \RPA & \RPA & \RFA & \RPA \\
    overflow-read-ptr-heap                   & \RPA & \RPA & \RPA & \RPA & \RPA & \RFA & \RPA \\
    overflow-read-ptr-data                   & \RPA & \RPA & \RPA & \RPA & \RPA & \RFA & \RPA \\
\bottomrule
  \end{tabular}
\end{center}

\begin{center}
  \footnotesize
  \begin{tabular}{lccccccc}
    \toprule
    & \rot{Default} & \rot{RELRO} & \rot{\shortstack[l]{Stack protection}} & \rot{CFI} &  \rot{All} & \rot{ASan} & \rot{None}\\
    \midrule
    overflow-write-cross-object-index-stack  & \RPA & \RPA & \RPA & \RPA & \RPA & \RNT & \RPA \\
    overflow-write-cross-object-index-heap   & \RPA & \RPA & \RPA & \RPA & \RPA & \RNT & \RPA \\
    overflow-write-cross-object-index-data   & \RPA & \RPA & \RPA & \RPA & \RPA & \RNT & \RPA \\
    overflow-write-cross-object-ptr-stack    & \RPA & \RPA & \RPA & \RPA & \RPA & \RNT & \RPA \\
    overflow-write-cross-object-ptr-heap     & \RPA & \RPA & \RPA & \RPA & \RPA & \RNT & \RPA \\
    overflow-write-cross-object-ptr-data     & \RPA & \RPA & \RPA & \RPA & \RPA & \RNT & \RPA \\
    underflow-write-cross-object-index-stack & \RPA & \RPA & \RPA & \RPA & \RPA & \RNT & \RPA \\
\bottomrule
  \end{tabular}
\end{center}

\newpage
\subsubsection{Temporal Safety}
\begin{center}
  \footnotesize
  \begin{tabular}{lccccccc}
    \toprule
    & \rot{Default} & \rot{RELRO} & \rot{\shortstack[l]{Stack protection}} & \rot{CFI} &  \rot{All} & \rot{ASan} & \rot{None}\\
    \midrule
    read-after-free-org-heap    & \RFA & \RFA & \RFA & \RFA & \RFA & \RFA & \RFA \\
    read-after-free-alias-heap  & \RFA & \RFA & \RFA & \RFA & \RFA & \RFA & \RFA \\
    write-after-free-heap       & \RPA & \RPA & \RPA & \RPA & \RPA & \RFA & \RPA \\
    \midrule
    reallocate-heap             & \RPA & \RPA & \RPA & \RPA & \RPA & \RFA & \RPA \\
    \midrule
    access-after-reclaim-heap   & \RFA & \RFA & \RFA & \RFA & \RFA & \RNT & \RFA \\
    write-before-reclaim-heap   & \RFA & \RFA & \RFA & \RFA & \RFA & \RNT & \RFA \\
    write-after-reclaim-heap    & \RPA & \RPA & \RPA & \RPA & \RPA & \RNT & \RPA \\
    \midrule
    read-after-free-alias-stack & \RPA & \RPA & \RPA & \RPA & \RPA & \RPA & \RPA \\
    write-after-free-stack      & \RPA & \RPA & \RPA & \RPA & \RPA & \RFA & \RPA \\
    \midrule
    reallocate-stack            & \RPA & \RPA &\RPA  & \RPA & \RPA & \RPA & \RPA \\
    \midrule
    access-after-reclaim-stack  & \RPA & \RPA & \RPA & \RPA & \RPA & \RPA & \RPA \\
    write-before-reclaim-stack  & \RPA & \RPA & \RPA & \RPA & \RPA & \RNT & \RPA \\
    write-after-reclaim-stack   & \RPA & \RPA & \RPA & \RPA & \RPA & \RPA & \RPA \\
    \bottomrule
  \end{tabular}
\end{center}

\subsubsection{Access Control}
\begin{center}
  \footnotesize
  \begin{tabular}{lccccccc}
    \toprule
    & \rot{Default} & \rot{RELRO} & \rot{\shortstack[l]{Stack protection}} & \rot{CFI} &  \rot{All} & \rot{ASan} & \rot{None}\\
    \midrule
    check-ASLR                  & \RPA & \RFA & \RPA & \RPA & \RFA & \RPA& \RPA \\
    read-func                   & \RPA & \RPA & \RPA & \RPA & \RPA & \RPA& \RPA \\
    read-GOT                    & \RPA & \RPA & \RPA & \RPA & \RPA & \RPA& \RPA \\
    \bottomrule
  \end{tabular}
\end{center}

\subsubsection{Pointer Integrity}

\begin{center}
  \footnotesize
  \begin{tabular}{lccccccc}
    \toprule
    & \rot{Default} & \rot{RELRO} & \rot{\shortstack[l]{Stack protection}} & \rot{CFI} &  \rot{All} & \rot{ASan} & \rot{None}\\
    \midrule
    func-pointer-assign     & \RPA & \RPA & \RPA & \RPA & \RPA & \RPA & \RPA \\
    func-pointer-arithmetic & \RFA & \RFA & \RFA & \RFA & \RFA & \RFA & \RFA \\
    \midrule
    read-vtable-pointer     & \RPA & \RPA & \RPA & \RPA & \RPA & \RFA & \RPA \\
    write-vtable-pointer    & \RPA & \RPA & \RPA & \RPA & \RPA & \RNT & \RPA \\
    \midrule
    modify-GOT              & \RPA & \REX & \RPA & \RPA & \REX & \RPA & \RPA \\
    \bottomrule
  \end{tabular}
\end{center}

\newpage
\subsubsection{Control-Flow Integrity}

\begin{center}
  \footnotesize
  \begin{tabular}{lccccccc}
    \toprule
    & \rot{Default} & \rot{RELRO} & \rot{\shortstack[l]{Stack protection}} & \rot{CFI} &  \rot{All} & \rot{ASan} & \rot{None}\\
    \midrule
    return-to-instruction-in-stack                   & \REX & \REX & \REX & \REX & \REX & \RNT & \RPA \\
    return-to-instruction-in-heap                    & \REX & \REX & \REX & \REX & \REX & \RNT & \RPA \\
    return-to-instruction-in-data                    & \REX & \REX & \REX & \REX & \REX & \RNT & \RPA \\
    return-to-instruction-in-rodata                  & \RPA & \RPA & \RPA & \RPA & \RPA & \RNT & \RPA \\
    \midrule
    return-to-wrong-call-site-within-static-analysis & \RPA & \RPA & \RPA & \RPA & \RPA & \RNT & \RPA \\
    return-to-wrong-call-site                        & \RPA & \RPA & \RPA & \RPA & \RPA & \RNT & \RPA \\
    return-to-non-call-site                          & \RPA & \RPA & \RPA & \RPA & \RPA & \RNT & \RPA \\
\bottomrule
  \end{tabular}
\end{center}

\end{document}